\pdfoutput=1
\documentclass[journal]{IEEEtran}

\usepackage{amsmath,amssymb,amsfonts}
\usepackage{algorithmic}
\usepackage{algorithm}
\usepackage{graphicx}
\usepackage{textcomp}
\usepackage{cite}
\usepackage{booktabs}
\usepackage{multirow}
\usepackage{bm}
\usepackage{xcolor}
\usepackage[hidelinks]{hyperref}
\usepackage{amsthm}
\usepackage{stfloats}
\usepackage{url}
\usepackage{acronym}
\usepackage{tikz}
\usetikzlibrary{arrows.meta}
\input{acronym}

\newcommand{\real}{\mathbb{R}}
\newcommand{\expect}{\mathbb{E}}
\newcommand{\complex}{\mathbb{C}}

\newcommand{\argmax}{\operatorname{arg\,max}}
\newcommand{\tr}{\operatorname{tr}}
\newcommand{\diag}{\operatorname{diag}}
\newcommand{\softmax}{\operatorname{softmax}}

\newtheorem{proposition}{Proposition}
\newtheorem{corollary}{Corollary}
\newtheorem{lemma}{Lemma}
\theoremstyle{definition}
\newtheorem{remark}{Remark}

\begin{document}

\title{Digital Twin-Conditioned RSS-Fingerprint Localization in Cell-Free Millimeter-Wave Networks}

\author{Varun~S.~Advani,~\IEEEmembership{Graduate~Student~Member,~IEEE,}
        Khalid~Kanaan,~\IEEEmembership{Graduate~Student~Member,~IEEE,}
        Ahmed~Nasser,~\IEEEmembership{Senior~Member,~IEEE,}
        Mohamed~Y.~Selim,~\IEEEmembership{Senior~Member,~IEEE,}
        and~Ahmed~M.~Eltawil,~\IEEEmembership{Senior~Member,~IEEE}%
\thanks{V. S. Advani and M. Y. Selim are with the Department of Electrical
and Computer Engineering, Iowa State University of Science and Technology,
Ames, IA 50011 USA (e-mail: vsadvani@iastate.edu; myoussef@iastate.edu).}%
\thanks{K. Kanaan, A. Nasser, and A. M. Eltawil are with the Computer, Electrical, and
Mathematical Sciences and Engineering Division, King Abdullah University
of Science and Technology (KAUST), Thuwal, KSA
(e-mail: khalid.kanaan@kaust.edu.sa; ahmednasser.ismail@kaust.edu.sa; ahmed.eltawil@kaust.edu.sa).}%
\vspace{-0.5em}}


\maketitle

\begin{abstract}
\Acp{DT} can generate dense synthetic fingerprints for localization via ray tracing, yet the sim-to-real domain gap limits their direct use, particularly in cell-free \ac{mmWave} deployments where distributed geometries and blockage-induced sparsity widen the mismatch.
We propose \ac{DTCond}, a framework that fuses real and \ac{DT} beam-domain fingerprints as complementary inputs through per-domain encoders, learnable per-\ac{AP} attention, and calibrated uncertainty.
We derive oracle and pipeline \acp{CRB} for \ac{DT}-aided fingerprint localization, proving that \ac{DT} information strictly tightens the positioning bound whenever spatial structure is preserved, with a heteroscedastic refinement that tightens the bound by a further 9.1\%.
On a real 28\,GHz cell-free testbed, the deployable two-stage pipeline, which reads the \ac{DT} fingerprint at a coarse \ac{kNN} estimate, achieves 0.284\,m mean error against 0.313\,m for real-only \acs{kNN}. With the \ac{DT} fingerprint read at the true position (oracle input), \ac{DT} conditioning achieves 0.186\,m mean error, validated through a ten-method ablation against \ac{DA} and transfer-learning baselines, and a 48\% improvement on unseen regions.
Multi-user self-attention provides 31--41\% per-user gain for emulated simultaneous users, and progressive calibration reduces the measurement budget by 50\%.
As a use case, localization drives network configuration in closed loop. Between beam sweeps, a \ac{DT} updated from measured \acs{RSS} and read at the tracked position estimate selects the 3 of 81 beam pairs that each \ac{AP} trains. With 49 calibration anchors in the update, the mean beam-selection loss stays below 1\,dB over 2\,m of motion, whereas reusing the last sweep loses 2.7\,dB after 1\,m.
\end{abstract}


\acresetall

\section{Introduction}
\label{sec:intro}

\IEEEPARstart{C}{ell-free} massive multiple-input multiple-output (MIMO) networks eliminate traditional cell boundaries by jointly serving users through distributed access points (APs) across the coverage area~\cite{ngo2017, bjornson2020, ammar2022cellfree}.
\acused{MIMO}\acused{AP}
When operating at millimeter-wave (mmWave) frequencies, such networks exploit directional beamforming with large antenna arrays, offering both high throughput and fine-grained spatial resolution~\cite{shahmansoori2018}.
\acused{mmWave}
Accurate user localization is a prerequisite for location-aware services including beam management, resource allocation, and handover optimization, making it a critical enabler for next-generation cell-free deployments~\cite{bjornson2021book}.
Fingerprint-based localization that maps \ac{RSS}-only beam-domain patterns to positions is well suited to cell-free networks because it naturally exploits the spatial diversity of multiple distributed \acp{AP} without requiring inter-\ac{AP} time synchronization or phase coherence~\cite{torres2015fingerprint}.
However, building a fingerprint database requires dense physical measurements, a costly process that scales with the service area size and the number of \acp{AP}.

Ray-tracing \acp{DT} address this measurement burden because synthetic fingerprints can be generated at arbitrary spatial density without any physical data collection~\cite{alkhateeb2023, masaracchia2024dt}.
Yet a persistent \emph{domain gap} exists between \ac{DT} predictions and real measurements.
Imprecise material properties in the 3D environment model, unmodeled scattering from small objects, and hardware imperfections absent from simulation all contribute to systematic differences between simulated and real fingerprints.
In practice, per-\ac{AP} Pearson correlations between real and \ac{DT} beam-domain power profiles can be low, making naive \ac{DT}-based approaches, such as training on simulated data and deploying on real data, unreliable.
Standard \ac{DA} techniques (such as \ac{DANN}~\cite{ganin2016} and \ac{CORAL}~\cite{sun2016coral}), which assume a distributional shift between structurally similar domains, also underperform because the gap reflects a fundamental mismatch in observation processes, not a simple distribution shift (Section~\ref{sec:exp_ablation}).
Despite growing interest in \ac{DT}-aided wireless systems~\cite{simeone2024, liu2024drec, bonati2021colosseum, morais2024fingerprint}, existing work focuses predominantly on beam management and single-cell scenarios, often validated only in simulation.
Prior \ac{DT}-aided localization uses the \ac{DT} as a substitute fingerprint database, a training source, or a positional reference for channel charting~\cite{renaudin2018rtfp, gao2024locdt, mateos2026dtcharting, akgun2024ota}. We extend our work in~\cite{advani2026pimrc} with new pipeline and heteroscedastic \ac{CRB} analysis, multi-user localization, progressive calibration, localization-driven beam and serving-\ac{AP} selection, and an expanded evaluation.

In this paper, we propose \ac{DTCond}, a localization framework that treats the \ac{DT} fingerprint as a complementary conditioning input rather than a source for pre-training or domain alignment.
\acs{DTCond} processes real and \ac{DT} beam-domain fingerprints through separate per-domain encoders, fuses them via learnable per-\ac{AP} attention weights that allocate position-dependent precision across the distributed \acp{AP}, and outputs both a position estimate and calibrated per-coordinate uncertainty.
The framework is grounded in a \ac{CRB} analysis that quantifies the positioning benefit of \ac{DT} information and motivates the architectural choices, including per-\ac{AP} weighting and heteroscedastic output.
A twin-in-the-loop calibration strategy progressively collects real measurements at spatially diverse locations, enabling deployment with limited field data.
At run time, the deployable two-stage pipeline reads the \ac{DT} fingerprint at a coarse real-only \ac{kNN} estimate, whereas the oracle input is the \ac{DT} fingerprint at the true position.
The main contributions are as follows.
\begin{enumerate}
\item \emph{Theory.} We formulate a joint observation model for \ac{DT}-aided fingerprint localization and derive oracle, pipeline, and heteroscedastic \acp{CRB}, showing that \ac{DT} information strictly tightens the positioning bound whenever spatial structure is preserved. A first-order sensitivity analysis reveals that \ac{DT} information is most valuable along weak eigendirections of the real-only \ac{FIM}, and a per-\ac{AP} decomposition provides a marginal \ac{DT} utility score motivating \ac{DT}-driven \ac{AP} attention. The pipeline \ac{CRB} accounts for retrieval mismatch and moves monotonically from the real-only to the oracle bound as the coarse-stage error shrinks; the heteroscedastic refinement tightens the oracle bound further wherever the local per-AP domain-gap variance stays below its pooled value.
\item \emph{Design principle and architecture.} We propose the \ac{DTCond} framework, which fuses real and \ac{DT} beam-domain \ac{RSS} fingerprints through per-domain neural encoders and learnable \ac{AP} attention to produce position estimates with calibrated uncertainty. The deployable pipeline achieves 0.284\,m mean error against 0.313\,m for real-only \ac{kNN}. With the oracle input, a ten-method ablation shows that \ac{DT} conditioning achieves 0.186\,m mean error (a 41\% reduction over real-only \ac{kNN}) and that the architecture adds a statistically significant 12\% over concatenation of the same inputs. Because DT-driven AP attention can express the per-AP precision allocation of the heteroscedastic CRB, \acs{DTCond} is structurally distinct from fusion baselines that lack per-AP weighting. Evaluation under \ac{LOLO} demonstrates a 48\% reduction in mean error on unseen spatial regions over real-only \acs{kNN}, with precision-aware attention extending this to 51\%.
\item \emph{Multi-user extension.} We introduce multi-user localization via \ac{DT}-mediated cross-user self-attention, achieving a 31--41\% reduction in mean per-user error over the single-user baseline for emulated simultaneous users. The \ac{DT} provides a shared spatial context that raises the multi-user benefit 1.8- to 2.5-fold compared to real-only self-attention, enabling cross-user information exchange regardless of physical inter-user distance.
\item \emph{Closed-loop operation.} We develop a twin-in-the-loop progressive calibration framework with spatially diverse measurement selection that halves the required real measurements while exceeding the accuracy of real-only baselines at full data. Calibration anchors thus improve the localizer, and, as a use case, its position estimates close the loop between the network and the \ac{DT}. Between beam sweeps, the updated \ac{DT}, read at the tracked position estimate, selects the 3 of 81 beam pairs that each \ac{AP} trains. With 49 calibration anchors in the update, the mean beam-selection loss stays below 1\,dB over 2\,m of motion, against 2.7\,dB after 1\,m for the last sweep.
\end{enumerate}

\begin{table}[t]
\centering
\caption{Summary of Key Notation}
\label{tab:notation}
\footnotesize
\renewcommand{\arraystretch}{0.95}
\begin{tabular}{@{}c p{0.68\columnwidth}@{}}
\toprule
\textbf{Symbol} & \textbf{Description} \\
\midrule
\multicolumn{2}{@{}l}{\emph{System Model}} \\
$M$ & Number of distributed \acp{AP} \\
$N_{\mathrm{tx}},\, N_{\mathrm{rx}}$ & \ac{AP} / \ac{UE} antenna elements (TX / RX beams) \\
$N_{\mathrm{b}}$ & Beam pairs per \ac{AP} ($N_{\mathrm{tx}} \!\times\! N_{\mathrm{rx}}$) \\
$N_{\mathrm{p}},\, L,\, S$ & Meas.\ positions, trajectories, snapshots per position \\
$\mathcal{A} \subset \real^2$, $\bm{q} = [x,\,y]^{\top}$ & Service area; \ac{UE} 2D position \\
$P_{\mathrm{tx}}$ & Transmit power \\
$\bm{H}_m(\bm{q}) \!\in\! \complex^{N_{\mathrm{rx}} \times N_{\mathrm{tx}}}$ & Channel matrix from \ac{AP}~$m$ to \ac{UE} \\
$\bm{w}_{m,i},\, \bm{v}_{j}$, $\eta_{m,n}$ & TX/RX beamforming vectors; receiver noise \\
\midrule
\multicolumn{2}{@{}l}{\emph{Fingerprints and Domain Gap}} \\
$\bm{r}^{(m)} \!\in\! \real^{N_{\mathrm{b}}}$ & Real fingerprint from \ac{AP}~$m$ \\
$\bm{d}^{(m)} \!\in\! \real^{N_{\mathrm{b}}}$ & \ac{DT} fingerprint from \ac{AP}~$m$ \\
$\bm{r},\,\bm{d} \!\in\! \real^{MN_{\mathrm{b}}}$ & Full concatenated fingerprints \\
$\bm{\delta}^{(m)}(\bm{q})$ & Domain gap at \ac{AP}~$m$ \\
$\sigma_{\mathrm{r}}^2,\, \sigma_{\Delta}^2$ & Noise / domain gap variance \\
$\bm{I}_{N_{\mathrm{b}}},\, \bm{I}_{MN_{\mathrm{b}}}$ & Identity matrices of indicated dimension \\
$\rho_m$ & Real-\ac{DT} Pearson correlation at \ac{AP}~$m$ \\
\midrule
\multicolumn{2}{@{}l}{\emph{CRB and FIM}} \\
$\bm{J}_{\mathrm{r}}(\bm{q}),\,\bm{J}_{\mathrm{d}}(\bm{q}),\,\bm{J}_{\mathrm{r\!+\!d}}(\bm{q})$ & Fisher information matrices (FIMs): real-only, \ac{DT}-only, joint \\
$\bm{f}_{\mathrm{r}}(\bm{q}),\,\bm{f}_{\mathrm{d}}(\bm{q})$ & Noise-free real / \ac{DT} fingerprint functions \\
$\partial \bm{f}_{\mathrm{r}}(\bm{q})/\partial \bm{q}$ & Fingerprint Jacobian \\
$\bar{\gamma}_m$ & Jacobian cosine similarity at \ac{AP}~$m$ \\
$\hat{c}$ & Jacobian alignment factor \\
$\tr(\cdot)$ & Matrix trace \\
\midrule
\multicolumn{2}{@{}l}{\emph{DTCond Architecture}} \\
$f_{\bm{\theta}}$ & DTCond model with parameters $\bm{\theta}$ \\
$\phi_{\mathrm{r}},\,\phi_{\mathrm{d}}$ & Per-domain encoder networks \\
$g(\cdot),\, \psi(\cdot)$ & \ac{AP} attention / fusion networks \\
$\bm{\alpha} \!\in\! \real^M$ & Learned \ac{AP} attention weights \\
$d_{\mathrm{e}},\, d_{\mathrm{h}},\, d_{\mathrm{f}}$ & Encoder embedding / hidden / fusion output dim. \\
$d_{\mathrm{a}},\, d_{\mathrm{u}}$ & \ac{AP} attn.\ hidden / multi-user feature dim. \\
$p_{\mathrm{d}}$ & Dropout probability \\
$\hat{\bm{q}},\,\bm{\sigma}$ & Predicted position and uncertainty \\
\midrule
\multicolumn{2}{@{}l}{\emph{Calibration}} \\
$\mathcal{C}$ & Calibration set of real measurements \\
$N_{\mathrm{c}} = |\mathcal{C}|$ & Calibration budget \\
$K_{\mathrm{u}}$ & Number of simultaneously localized users \\
\bottomrule
\end{tabular}
\end{table}

Table~\ref{tab:notation} summarizes the key notation.

\begin{table*}[t]
\centering
\caption{Comparison with Representative Related Work on \acs{DT}-Aided Wireless Systems, Localization, and Cell-Free Positioning ($^{\dagger}$Users Emulated From Single-User Measurements)}
\label{tab:literature}
\footnotesize
\resizebox{\textwidth}{!}{%
\begin{tabular}{@{}lcccccccccl@{}}
\toprule
\textbf{Reference} & \textbf{\acs{DT}-Aided} & \textbf{Cell-Free} & \textbf{RSS-Only} & \textbf{Localization} & \textbf{\acs{CRB}} & \textbf{DA} & \textbf{Multi-User} & \textbf{Prog.\ Calib.} & \textbf{Real Data} & \textbf{Key Focus} \\
\midrule
Alkhateeb \emph{et al.} \cite{alkhateeb2023}    & \checkmark &            &            &            &            &            &            &            & \checkmark & Beam management \\
Charan \emph{et al.} \cite{charan2024}           & \checkmark &            &            &            &            &            &            &            & \checkmark & Vision-aided beams \\
Ruah \emph{et al.} \cite{simeone2024}            & \checkmark &            &            &            &            &            &            &            &            & Bayesian \acs{DT} framework \\
Cui \emph{et al.} \cite{liu2024drec}             & \checkmark & \checkmark &            &            &            &            &            &            &            & \acs{RIS} cell-free management \\
Morais \& Alkhateeb \cite{morais2024fingerprint}  & \checkmark &            & \checkmark & \checkmark &            &            &            &            &            & Massive fingerprinting \\
Demirhan \& Alkhateeb \cite{demirhan2023transfer} & \checkmark &           &            &            &            &            &            &            & \checkmark & Transfer learning \\
\midrule
Renaudin \emph{et al.} \cite{renaudin2018rtfp} & \checkmark &  & \checkmark & \checkmark & &   &            &            & \checkmark & Ray-traced fingerprints \\
Gao \emph{et al.} \cite{gao2024locdt}    & \checkmark &        &            & \checkmark & &            &            &            &            & Localization-oriented \acs{DT} \\
Akgun \emph{et al.} \cite{akgun2024ota}  & \checkmark &        & \checkmark & \checkmark & &   &            &            & \checkmark & Testbed \acs{DT} positioning \\
Saribekyan \emph{et al.} \cite{saribekyan2026sim2real} & \checkmark & & \checkmark & \checkmark & & \checkmark & & & \checkmark & Sim-to-real \acs{RSSI} positioning \\
\midrule
Shahmansoori \emph{et al.} \cite{shahmansoori2018}&           &            &            & \checkmark & \checkmark &            &            &            &            & \acs{mmWave} \acs{ToA}/\acs{AoA} \\
Shen \& Win \cite{shen2010}                       &           &            &            & \checkmark & \checkmark &            &            &            &            & Wideband \acs{FIM} \\
He \& Chan \cite{torres2015fingerprint}            &           &            & \checkmark & \checkmark &            &            &            &            & \checkmark & Wi-Fi fingerprinting \\
Gante \emph{et al.} \cite{gante2020deepfp}        &           &            &            & \checkmark &            &            &            &            &            & \acs{DL} \acs{mmWave} positioning \\
Advani \emph{et al.} \cite{11431957}               &           &            & \checkmark & \checkmark &            &            &            &            & \checkmark & \acs{RIS}-aided \acs{RSSI} loc. \\
\midrule
Shen \emph{et al.} \cite{kumar2025cellfree}       &           & \checkmark &            & \checkmark &            &            &            &            &            & Cell-free fingerprint loc. \\
Savic \& Larsson \cite{denis2020cellfree_loc}     &           & \checkmark & \checkmark & \checkmark &            &            &            &            &            & Distributed \acs{MIMO} fingerprint \\
\midrule
\textbf{This work}                                & \checkmark & \checkmark & \checkmark & \checkmark & \checkmark & \checkmark & \checkmark$^{\dagger}$ & \checkmark & \checkmark & \textbf{DT-cond.\ RSS loc.} \\
\bottomrule
\end{tabular}%
}
\end{table*}

\section{Related Work}
\label{sec:related}

\subsection{Digital Twins for Wireless Networks}
\label{sec:related_dt}

\Acp{DT} have become widely adopted for wireless network design~\cite{wu2021dt6g, masaracchia2024dt}, with Alkhateeb and coworkers pioneering \ac{DT}-aided beam management via ray tracing~\cite{alkhateeb2019deepmimo, alkhateeb2023, charan2024} and transfer learning~\cite{demirhan2023transfer}.
Other applications include Bayesian calibration~\cite{simeone2024}, \ac{RIS} management~\cite{liu2024drec}, and network experimentation~\cite{bonati2021colosseum}, but existing work focuses on beam management or resource allocation in single-cell settings with few real-data evaluations.
\looseness=-1 Site-specific ray tracing forms the radio layer of such twins~\cite{alkhateeb2019deepmimo, hoydis2023sionna}, runs inside network simulators~\cite{pegurri2025ns3}, is calibrated from measurements~\cite{hoydis2024radio}, and has guided integrated sensing and communication~\cite{jiang2024dtisac}. Closed-loop twins account for their own uncertainty~\cite{simeone2024} or are updated when they diverge from reality~\cite{salehi2026darwin}. For localization, ray-traced databases have replaced surveyed fingerprints~\cite{renaudin2018rtfp, morais2024fingerprint}, and twins have trained positioning networks~\cite{gao2024locdt}, provided positional references for channel charts~\cite{mateos2026dtcharting}, and supported positioning on a 5G testbed~\cite{akgun2024ota} (Table~\ref{tab:literature}). \acs{DTCond} instead pairs measured and \ac{DT} fingerprints and relates the gain to the domain gap through the \ac{CRB}.

\subsection{Localization in Cell-Free and \acs{mmWave} Systems}
\label{sec:related_fingerprint}
RSS-only fingerprint-based localization maps received power patterns to positions~\cite{torres2015fingerprint}, with \ac{DL} approaches using convolutional~\cite{wang2017csi, arnold2019cnnloc} and \ac{mmWave}-specific architectures~\cite{gante2020deepfp, vieira2017deep}.
Our prior work~\cite{11431957} demonstrated RSS-based \ac{mmWave} localization in an \ac{RIS}-aided \ac{NLoS} environment (0.25\,m on a real 28\,GHz testbed); the present work extends to cell-free multi-\ac{AP} deployments with \ac{DT} conditioning and \ac{CRB} analysis.
Model-based \ac{mmWave} positioning exploits \ac{ToA} and \ac{AoA}~\cite{shahmansoori2018, gustafsson2005positioning}.
Cell-free and distributed-MIMO localization has been studied via RSS fingerprinting in distributed massive \ac{MIMO}~\cite{denis2020cellfree_loc}, model-based \acs{mmWave} positioning with inter-\ac{AP} synchronization~\cite{wymeersch2018cellfree_loc}, and \acs{mmWave} cell-free fingerprint schemes~\cite{kumar2025cellfree}, but all rely on simulation; the user-centric paradigm~\cite{ammar2022cellfree} motivates localization-aware \ac{AP} selection, which Section~\ref{sec:calibration} realizes through the \ac{DT}.

\subsection{\texorpdfstring{\ac{DA}}{DA} for Wireless Systems}
\label{sec:related_da}

The sim-to-real gap has motivated several alignment-based approaches: \ac{DANN}~\cite{ganin2016} learns domain-invariant features through adversarial training, \ac{CORAL}~\cite{sun2016coral} aligns second-order statistics, and \ac{MMD} matches distributions through kernel mean embeddings~\cite{gretton2012mmd}.
Transfer learning has been applied to radar-aided beam prediction~\cite{demirhan2023transfer}.
For sim-to-real \ac{RSSI} positioning, matching the simulated and measured distributions mattered more than physical realism~\cite{saribekyan2026sim2real}.
All these methods assume a removable distributional shift between structurally similar source and target domains; however, this assumption does not hold for \ac{DT}-aided localization.
The real and \ac{DT} fingerprints are produced by different observation processes, namely stochastic measurement and deterministic ray tracing, and forcing distributional alignment removes the spatial structure that the \ac{DT} contributes.
Section~\ref{sec:exp_ablation} shows that \ac{DA} baselines underperform real-only methods for this reason.

\subsection{Theoretical Bounds for Localization}
\label{sec:related_crb}

The \ac{CRB} lower-bounds the variance of any unbiased estimator via the inverse \ac{FIM}.
Foundational frameworks for wideband~\cite{shen2010} and cooperative~\cite{win2018crb, win2011cooperative} localization show per-anchor \ac{FIM} contributions; extensions cover \ac{RSS}-based bounds~\cite{li2007rss_crb} and multi-measurement types~\cite{gustafsson2005positioning}.
The misspecified \ac{CRB}~\cite{richmond2015misspecified} extends these results when the assumed observation distribution is incorrect.
The joint observation model in which a noisy real fingerprint and a deterministic \ac{DT} fingerprint are available at the same position lies outside these frameworks and is analyzed in Section~\ref{sec:system_crb}.

\section{System Model}
\label{sec:system}

\begin{figure}[!t]
\centering
\begin{tikzpicture}
\node[anchor=south west, inner sep=0] (img) at (0,0) {\includegraphics[width=\columnwidth]{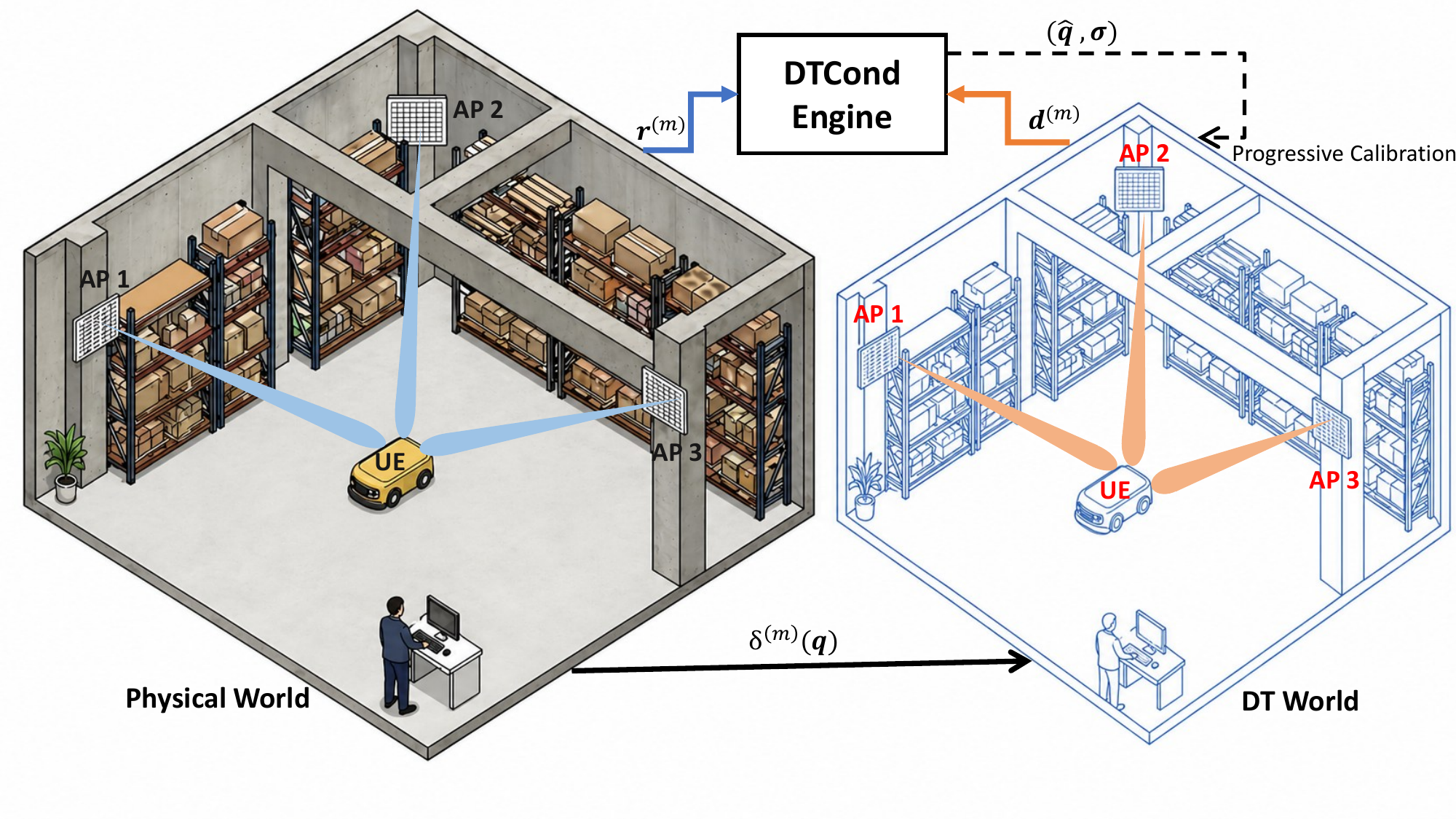}};
\begin{scope}[x={(img.south east)}, y={(img.north west)}]
\draw[-{To[length=1.4mm, width=2.2mm]}, line width=0.75pt] (0.695,0.100) -- (0.415,0.100);
\node[font=\scriptsize] at (0.555,0.058) {$\mathcal{B}_m$};
\end{scope}
\end{tikzpicture}
\caption{Twin-in-the-loop system architecture. Distributed \acp{AP} collect beam-domain fingerprints $\bm{r}^{(m)}$; a ray-tracing \ac{DT} generates synthetic fingerprints $\bm{d}^{(m)}$. DTCond fuses both with progressive calibration. Localization closes the loop. The \ac{DT} is read at the position estimate $(\hat{\bm{q}}, \bm{\sigma})$ (dashed arrow) and updated by the gap $\bm{\delta}^{(m)}(\bm{q})$ estimated from measured \ac{RSS} (upper solid arrow). It selects the beam pairs $\mathcal{B}_m$ that the \acp{AP} train (lower arrow).}
\label{fig:system}
\end{figure}

This section presents the cell-free \ac{mmWave} network model, defines the \ac{DT} and domain gap, and formulates the localization problem.
The twin-in-the-loop system architecture is illustrated in Fig.~\ref{fig:system}.

\subsection{Cell-Free \acs{mmWave} Network Model}
\label{sec:system_cellfree}

We consider a cell-free \ac{mmWave} network comprising $M$ distributed \acp{AP} that jointly serve \acfp{UE} within a service area $\mathcal{A} \subset \mathbb{R}^2$.
Each \ac{AP}~$m \in \{1, \ldots, M\}$ is equipped with a \ac{UPA} of $N_{\mathrm{tx}}$ antenna elements and employs a \ac{DFT} codebook of $N_{\mathrm{tx}}$ directional transmit beams, assuming a non-oversampled codebook. 
The \ac{UE} is equipped with a \ac{UPA} of $N_{\mathrm{rx}}$ antenna elements using a corresponding \ac{DFT} codebook of $N_{\mathrm{rx}}$ receive beams.
During beam sweeping, \ac{AP}~$m$ transmits sequentially across all transmit beams while the \ac{UE} scans all receive beams, yielding $N_{\mathrm{b}} = N_{\mathrm{tx}} \times N_{\mathrm{rx}}$ beam-pair power measurements per \ac{AP}.
Both the \ac{AP} and the \ac{UE} apply beamforming, at the transmitter and at the receiver, with fixed codebooks that require no channel state information, unlike channel-adaptive conjugate beamforming. Localization uses the received powers of the full beam sweep, whichever beam pair the network then selects for communication.

For a \ac{UE} at unknown position $\bm{q} = [x,\, y]^{\top} \in \mathcal{A}$, the received power under beam pair $n \in \{1, \ldots, N_{\mathrm{b}}\}$ from \ac{AP}~$m$ is
\begin{equation}
\label{eq:rx_power}
P_{m,n}(\bm{q}) = P_{\mathrm{tx}} \left| \bm{v}_{j(n)}^{H} \bm{H}_m(\bm{q})\, \bm{w}_{m,i(n)} \right|^2 + \eta_{m,n},
\end{equation}
where $P_{\mathrm{tx}}$ is the transmit power, $\bm{H}_m(\bm{q}) \in \mathbb{C}^{N_{\mathrm{rx}} \times N_{\mathrm{tx}}}$ is the position-dependent \ac{MIMO} channel matrix from \ac{AP}~$m$ to the \ac{UE}, $\bm{w}_{m,i(n)} \in \mathbb{C}^{N_{\mathrm{tx}}}$ and $\bm{v}_{j(n)} \in \mathbb{C}^{N_{\mathrm{rx}}}$ are the transmit and receive beamforming vectors for beam pair~$n$ (with $n \mapsto (i(n), j(n))$ the bijection from $\{1, \ldots, N_{\mathrm{b}}\}$ to $\{1, \ldots, N_{\mathrm{tx}}\} \times \{1, \ldots, N_{\mathrm{rx}}\}$), and $\eta_{m,n}$ denotes the receiver noise contribution.
The beam-domain fingerprint from \ac{AP}~$m$, measured at the \ac{UE}, collects the received powers across all $N_{\mathrm{b}}$ beam pairs
\begin{equation}
\label{eq:fp_ap}
\bm{r}^{(m)}(\bm{q}) = \left[P_{m,1}(\bm{q}),\, \ldots,\, P_{m,N_{\mathrm{b}}}(\bm{q})\right]^{\top} \in \mathbb{R}^{N_{\mathrm{b}}}.
\end{equation}
The full cell-free fingerprint stacks those of all $M$ \acp{AP}
\begin{equation}
\label{eq:fp_full}
\bm{r}(\bm{q}) = \left[\bm{r}^{(1)\top}\!(\bm{q}),\; \ldots,\; \bm{r}^{(M)\top}\!(\bm{q})\right]^{\top} \in \mathbb{R}^{MN_{\mathrm{b}}}.
\end{equation}
The cell-free deployment enriches the spatial discriminability of $\bm{r}(\bm{q})$ by providing complementary angular views from geographically separated \acp{AP}.

\subsection{Digital Twin Model and Domain Gap}
\label{sec:system_dt}

A \ac{DT} of the wireless environment generates synthetic fingerprints via a deterministic ray-tracing simulation.
The \ac{DT} produces a simulated beam-domain fingerprint $\bm{d}^{(m)}(\bm{q}) \in \mathbb{R}^{N_{\mathrm{b}}}$ $\forall (\bm{q}, m) \in \mathcal{A} \times \{1, \ldots, M\}$, and the full \ac{DT} fingerprint is $\bm{d}(\bm{q}) = [\bm{d}^{(1)\top}\!(\bm{q}),\, \ldots,\, \bm{d}^{(M)\top}\!(\bm{q})]^{\top} \in \mathbb{R}^{MN_{\mathrm{b}}}$.
Unlike real measurements, \ac{DT} fingerprints are deterministic and can be generated at arbitrary spatial density without measurements.

A systematic domain gap exists between real and \ac{DT} fingerprints.
All fingerprints are converted to the dBm domain via elementwise $10\log_{10}(\cdot)$ before further processing; for notational compactness, $\bm{r}^{(m)}$ and $\bm{d}^{(m)}$ denote dBm-domain quantities from this point onward.
In dBm, multiplicative power differences become additive, and the real fingerprint decomposes as
\begin{equation}
\label{eq:domain_gap}
\bm{r}^{(m)}(\bm{q}) = \bm{d}^{(m)}(\bm{q}) + \bm{\delta}^{(m)}(\bm{q}) + \bm{n}^{(m)},
\end{equation}
where $\bm{\delta}^{(m)}(\bm{q}) \in \mathbb{R}^{N_{\mathrm{b}}}$ is the position- and \ac{AP}-dependent domain gap capturing systematic differences between the \ac{DT} and reality (imprecise material properties, unmodeled scattering, and hardware imperfections), and $\bm{n}^{(m)} \sim \mathcal{N}(\bm{0},\, \sigma_{\mathrm{r}}^2 \bm{I}_{N_{\mathrm{b}}})$ is the measurement noise.
This additive decomposition is exact in dBm, where the noise $\bm{n}^{(m)}$ represents the snapshot-to-snapshot variation after dBm conversion. The Gaussian approximation is validated empirically in Section~\ref{sec:exp_crb}.
The domain gap $\bm{\delta}^{(m)}(\bm{q})$ is spatially varying, generally nonzero-mean, and can be large in magnitude.
We quantify it through the per-\ac{AP} Pearson correlation
\begin{equation}
\label{eq:correlation}
\rho_m = \frac{\sum_{k=1}^{N_{\mathrm{p}}} \left(\bm{r}_k^{(m)} - \bar{\bm{r}}^{(m)}\right)^{\top}\!\left(\bm{d}_k^{(m)} - \bar{\bm{d}}^{(m)}\right)}
{\sqrt{\sum_{k=1}^{N_{\mathrm{p}}}\left\|\bm{r}_k^{(m)} - \bar{\bm{r}}^{(m)}\right\|^2}\;\sqrt{\sum_{k=1}^{N_{\mathrm{p}}}\left\|\bm{d}_k^{(m)} - \bar{\bm{d}}^{(m)}\right\|^2}},
\end{equation}
where $k \in \{1, \ldots, N_{\mathrm{p}}\}$ indexes the measurement positions, $\bar{\bm{r}}^{(m)} = \frac{1}{N_{\mathrm{p}}}\sum_{k=1}^{N_{\mathrm{p}}} \bm{r}_k^{(m)}$ is the sample mean of the real fingerprints from \ac{AP}~$m$, and $\bar{\bm{d}}^{(m)} = \frac{1}{N_{\mathrm{p}}}\sum_{k=1}^{N_{\mathrm{p}}} \bm{d}_k^{(m)}$ is the corresponding \ac{DT} sample mean.
Low values of $\rho_m$ indicate large domain gaps, but do not preclude the \ac{DT} from preserving \emph{spatial structure}, a key distinction we formalize next.

\begin{figure}[!t]
\centering
\includegraphics[width=0.78\columnwidth]{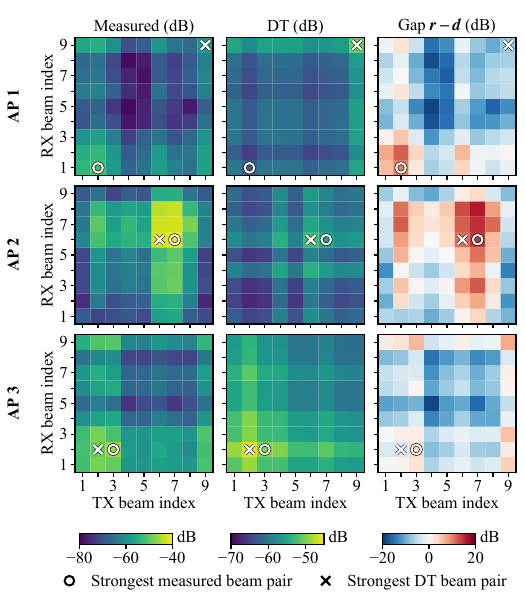}
\caption{Measured \ac{RSS} $\bm{r}^{(m)}$ and \ac{DT} prediction $\bm{d}^{(m)}$, both power-balanced (Section~\ref{sec:system_preprocessing}), and their difference (columns) for \acp{AP}~1--3 (rows) over the $9 \times 9$ transmit (horizontal) and receive (vertical) beams at $(0.9, 3.6)$\,m, the star in Fig.~\ref{fig:rss_maps}. Circles and crosses mark the strongest measured and \ac{DT} beam pairs. They lie within one index at \acp{AP}~2 and~3 despite the level differences and differ at \ac{AP}~1 ($\rho_1 = 0.17$).}
\label{fig:beam_heatmaps}
\end{figure}

\begin{remark}[Spatial fidelity versus absolute fidelity]
\label{rem:spatial}
The domain gap degrades absolute fingerprint accuracy, but the \ac{DT} can preserve the \emph{spatial structure} of the fingerprint field. Specifically, the spatial gradient $\partial \bm{d}^{(m)}(\bm{q})/\partial \bm{q}$ can be well correlated with $\partial \bm{r}^{(m)}(\bm{q})/\partial \bm{q}$ even when $\|\bm{\delta}^{(m)}(\bm{q})\|$ is large, as illustrated in Fig.~\ref{fig:beam_heatmaps}.
Physically, spatial structure is how the fingerprint changes as the \ac{UE} moves. In $\bm{\delta}^{(m)}(\bm{q}) = \bar{\bm{\delta}}^{(m)} + \tilde{\bm{\delta}}^{(m)}(\bm{q})$, the spatial mean $\bar{\bm{\delta}}^{(m)}$ biases every absolute power but $\partial \bar{\bm{\delta}}^{(m)}/\partial \bm{q} = \bm{0}$, so only $\tilde{\bm{\delta}}^{(m)}$ enters Lemma~\ref{lem:sys_gradient}.
Proposition~\ref{prop:crb} formalizes this distinction.
We quantify spatial gradient preservation via the per-\ac{AP} Jacobian cosine similarity
\begin{equation}
\label{eq:jacobian_cosine}
\bar{\gamma}_m = \frac{\bigl\langle \partial \bm{r}^{(m)}/\partial \bm{q},\, \partial \bm{d}^{(m)}/\partial \bm{q} \bigr\rangle_F}{\|\partial \bm{r}^{(m)}/\partial \bm{q}\|_F \,\|\partial \bm{d}^{(m)}/\partial \bm{q}\|_F},
\end{equation}
which measures how well the \ac{DT}'s spatial variation tracks that of the real measurements at each \ac{AP}; the empirical values reported in Section~\ref{sec:exp_crb} provide direct confirmation.
In particular, a \ac{DT} with large systematic bias but rich spatial variation still contributes substantial positioning information, whereas one with $\partial \bm{d}(\bm{q})/\partial \bm{q} \approx \bm{0} \in \mathbb{R}^{MN_{\mathrm{b}} \times 2}$ provides no benefit regardless of calibration, motivating an architecture that exploits spatial structure rather than absolute power.
\end{remark}

\begin{lemma}[Gradient preservation implies alignment]
\label{lem:sys_gradient}
Let $\bm{D} = \partial \bm{d}^{(m)}/\partial \bm{q}$ and $\bm{E} = \partial \tilde{\bm{\delta}}^{(m)}/\partial \bm{q}$ at $\bm{q}$, so that the noise-free real Jacobian is $\bm{D} + \bm{E}$. If $\bm{D} \neq \bm{0}$ and $\kappa_m = \|\bm{E}\|_F / \|\bm{D}\|_F < 1$, the cosine~\eqref{eq:jacobian_cosine} between $\bm{D}$ and $\bm{D} + \bm{E}$, $\cos\theta$, satisfies $\cos\theta \geq \sqrt{1 - \kappa_m^2}$ $\forall \bar{\bm{\delta}}^{(m)} \in \mathbb{R}^{N_{\mathrm{b}}}$, hence $\forall \|\bm{\delta}^{(m)}(\bm{q})\| \geq 0$, with equality $\Leftrightarrow \bm{E} \perp (\bm{D} + \bm{E})$.
\end{lemma}
\begin{proof}
The angle $\theta$ between $\bm{D}$ and $\bm{D} + \bm{E}$ satisfies $\theta < \pi/2$, since $\langle \bm{D}, \bm{D} + \bm{E}\rangle_F \geq \|\bm{D}\|_F^2 - \|\bm{D}\|_F \|\bm{E}\|_F = (1 - \kappa_m)\|\bm{D}\|_F^2 > 0$ by the Cauchy--Schwarz inequality, and the distance from $\bm{D}$ to the line $\{t(\bm{D} + \bm{E})\}_{t \in \mathbb{R}}$ is $\|\bm{D}\|_F \sin\theta = \min_{t} \|\bm{D} - t(\bm{D} + \bm{E})\|_F \leq \|\bm{D} - (\bm{D} + \bm{E})\|_F = \kappa_m \|\bm{D}\|_F$, with equality $\Leftrightarrow$ $t = 1$ is the minimizer, i.e., $\bm{E} \perp (\bm{D} + \bm{E})$. Hence $\cos\theta = \sqrt{1 - \sin^2\theta} \geq \sqrt{1 - \kappa_m^2}$.
\end{proof}

\looseness=-1 \noindent\emph{Examples.} A constant offset, such as cable loss or a transmit-power error, gives $\kappa_m = 0$. At 5\,m from an \ac{AP}, a path-loss exponent error of 0.3 adds a gap gradient of $10 \cdot 0.3/(5 \ln 10) \approx 0.26$\,dB/m. Modeling concrete (relative permittivity $5.24 - j0.40$ at 28\,GHz) as $4.0 - j0.30$ biases a reflection by 0.8--1.4\,dB but its gradient by $\leq 0.25$\,dB/m, since reflection loss follows the slowly varying incidence angle. Both lie far below the 2.3--7.0\,dB/m beam-pattern gradients, so each of these errors alone gives $\kappa_m \ll 1$. An \ac{AP} position error $\Delta$ gives $\kappa_m \approx \Delta / w_{\mathrm{fp}}$ for the beam footprint $w_{\mathrm{fp}} \in [1.0, 2.3]$\,m. An unmodeled blocker violates $\kappa_m < 1$, as its shadow edge carries gap gradients near 75\,dB/m.

\subsection{Problem Formulation}
\label{sec:system_problem}

Given a calibration set $\mathcal{C} = \{(\tilde{\bm{r}}_k, \tilde{\bm{d}}_k, \bm{q}_k)\}_{k=1}^{N_{\mathrm{c}}}$ of $N_{\mathrm{c}}$ paired real and \ac{DT} fingerprints at known positions, and a \ac{DT} model that can generate fingerprints over $\mathcal{A}$ via ray tracing, we seek a localization function $f_{\bm{\theta}}\!: \mathbb{R}^{MN_{\mathrm{b}}} \times \mathbb{R}^{MN_{\mathrm{b}}} \to \mathbb{R}^2 \times \mathbb{R}_{>0}^2$ that maps a real fingerprint $\tilde{\bm{r}}_k$ and its corresponding \ac{DT} fingerprint $\tilde{\bm{d}}_k$ to a position estimate $\hat{\bm{q}}_k$ and uncertainty estimate $\bm{\sigma}_k$
\begin{equation}
\label{eq:problem}
(\hat{\bm{q}}_k,\, \bm{\sigma}_k) = f_{\bm{\theta}}(\tilde{\bm{r}}_k,\, \tilde{\bm{d}}_k).
\end{equation}
At inference, $f_{\bm{\theta}}$ is applied to any new fingerprint pair $(\tilde{\bm{r}},\, \tilde{\bm{d}})$, including positions $\bm{q} \notin \{\bm{q}_k\}_{k=1}^{N_{\mathrm{c}}}$.
The learning objective over the calibration set is
\begin{equation}
\label{eq:objective}
\min_{\bm{\theta}} \; \frac{1}{N_{\mathrm{c}}} \sum_{k=1}^{N_{\mathrm{c}}} \mathcal{L}\!\left(f_{\bm{\theta}}(\tilde{\bm{r}}_k, \tilde{\bm{d}}_k);\; \bm{q}_k\right),
\end{equation}
where $\mathcal{L}$ transitions from \ac{MSE} on $\hat{\bm{q}}_k$ to heteroscedastic \ac{NLL} on $(\hat{\bm{q}}_k, \bm{\sigma}_k)$ during training (Section~\ref{sec:framework_training}).
Two challenges distinguish this from standard fingerprint localization: (i)~the \ac{DT} input $\tilde{\bm{d}}$ lies in a different statistical domain from $\tilde{\bm{r}}$, requiring per-domain processing; (ii)~the limited budget $N_{\mathrm{c}}$ motivates a progressive, information-maximizing strategy (Section~\ref{sec:calibration}).

\section{Cram\'{e}r--Rao Bound for \acs{DT}-Aided Localization}
\label{sec:system_crb}

We now derive \acp{CRB} to quantify the benefit of \ac{DT} information. 
The oracle \ac{CRB} assumes access to the \ac{DT} fingerprint at the true position; its role is explanatory, isolating how the \ac{DT}'s spatial variation enters the localization information.
The pipeline \ac{CRB} of Proposition~\ref{prop:pipeline_crb} characterizes the deployable two-stage pipeline of Section~\ref{sec:inference}.

Let $\bm{f}_{\mathrm{r}}(\bm{q}), \bm{f}_{\mathrm{d}}(\bm{q}), \bm{\delta}(\bm{q}) \in \mathbb{R}^{MN_{\mathrm{b}}}$ denote the noise-free real fingerprint, concatenated \ac{DT} fingerprint, and concatenated domain gap at position~$\bm{q}$, where $\bm{f}_{\mathrm{d}}(\bm{q}) = [\bm{d}^{(1)\top}\!(\bm{q}), \ldots, \bm{d}^{(M)\top}\!(\bm{q})]^{\top}$ and $\bm{\delta}(\bm{q})$ is defined analogously. Concatenating~\eqref{eq:domain_gap} across all \acp{AP} gives $\bm{f}_{\mathrm{r}}(\bm{q}) = \bm{f}_{\mathrm{d}}(\bm{q}) + \bm{\delta}(\bm{q})$, so the real measurement follows the observation model

\begin{equation}
\label{eq:obs_real}
\bm{r}(\bm{q}) = \bm{f}_{\mathrm{r}}(\bm{q}) + \bm{n}, \quad \bm{n} \sim \mathcal{N}(\bm{0},\, \sigma_{\mathrm{r}}^2 \bm{I}_{MN_{\mathrm{b}}}),
\end{equation}
where $\bm{n} = [\bm{n}^{(1)\top}\!, \ldots, \bm{n}^{(M)\top}]^{\top}$ concatenates the per-\ac{AP} noise vectors, with $\bm{n}^{(m)} \perp\!\!\!\perp \bm{n}^{(m')}$ $\forall m \neq m'$, and $\sigma_{\mathrm{r}}^2$ is the per-element measurement noise variance, estimated from the $S$ repeated snapshots as $\hat{\sigma}_{\mathrm{r}}^2 = \frac{1}{N_{\mathrm{p}} M N_{\mathrm{b}}} \sum_{k,m,n} \mathrm{Var}_s\!\left[10\log_{10}P_{m,n,s}(\bm{q}_k)\right]$, where $\mathrm{Var}_s[\cdot]$ denotes the sample variance over the $S$ snapshots.

The \ac{FIM} for estimating $\bm{q}$ from $\bm{r}$ alone is~\cite{shen2010}
\begin{equation}
\label{eq:fim_real}
\bm{J}_{\mathrm{r}}(\bm{q}) = \frac{1}{\sigma_{\mathrm{r}}^2}
\left(\frac{\partial \bm{f}_{\mathrm{r}}(\bm{q})}{\partial \bm{q}}\right)^{\!\!\top}
\frac{\partial \bm{f}_{\mathrm{r}}(\bm{q})}{\partial \bm{q}} \in \mathbb{R}^{2 \times 2},
\end{equation}
where $\partial \bm{f}_{\mathrm{r}}(\bm{q})/\partial \bm{q} \in \mathbb{R}^{MN_{\mathrm{b}} \times 2}$ is the fingerprint Jacobian with respect to the two-dimensional position $\bm{q} = [x,y]^{\top}$.

In practice, this Jacobian is estimated numerically from local spatial neighborhoods (Section~\ref{sec:exp_crb}), with rows corresponding to \ac{AP}~$m$ encoding that \ac{AP}'s spatial sensitivity.

To quantify the maximum \ac{DT} information, we treat the \ac{DT} fingerprint as a matched auxiliary observation.
Let $\bm{f}_{\mathrm{d}}(\bm{q}) \in \mathbb{R}^{MN_{\mathrm{b}}}$ denote the \ac{DT} fingerprint function, subject to effective per-element noise variance $\sigma_{\Delta}^2$---the per-element variance of $\bm{r} - \bm{d}$ across positions, computed on the snapshot-averaged fingerprints (subscript $\Delta$ distinguishes it from $\sigma_{\mathrm{r}}^2$).
Treating $\sigma_{\Delta}^2$ as a single scalar (identical $\forall (\bm{q}, m) \in \mathcal{A} \times \{1, \ldots, M\}$) is a simplification that Proposition~\ref{prop:het_crb} refines per \ac{AP}.

The \ac{DT} contribution to the \ac{FIM} is
\begin{equation}
\label{eq:fim_dt}
\bm{J}_{\mathrm{d}}(\bm{q}) = \frac{1}{\sigma_{\Delta}^2}
\left(\frac{\partial \bm{f}_{\mathrm{d}}(\bm{q})}{\partial \bm{q}}\right)^{\!\!\top}
\frac{\partial \bm{f}_{\mathrm{d}}(\bm{q})}{\partial \bm{q}}.
\end{equation}
Since real measurement noise and \ac{DT} modeling errors are conditionally independent (arising from physically distinct processes), the joint \ac{FIM} is additive:
\begin{equation}
\label{eq:fim_joint}
\bm{J}_{\mathrm{r\!+\!d}}(\bm{q}) = \bm{J}_{\mathrm{r}}(\bm{q}) + \bm{J}_{\mathrm{d}}(\bm{q}).
\end{equation}
The \ac{CRB} on the root-mean-square positioning error at position $\bm{q}$ is
\begin{equation}
\label{eq:crb}
\mathrm{CRB}_{\star}(\bm{q}) = \sqrt{\tr\!\left(\bm{J}_{\star}^{-1}(\bm{q})\right)}, \quad \star \in \{\mathrm{r},\, \mathrm{r\!+\!d}\}.
\end{equation}
The joint $\mathrm{CRB}_{\mathrm{r\!+\!d}}(\bm{q})$ is an oracle benchmark assuming access to the \ac{DT} fingerprint at the true position; the practical retrieved-\ac{DT} pipeline is evaluated in Section~\ref{sec:exp_ablation}.

\begin{proposition}[Oracle monotonicity of DT information]
\label{prop:crb}
Let $\bm{q} \in \mathcal{A}$ satisfy $\bm{J}_{\mathrm{r}}(\bm{q}) \succ \bm{0}$. Then
\begin{equation}
\label{eq:crb_inequality}
\mathrm{CRB}_{\mathrm{r\!+\!d}}(\bm{q}) \leq \mathrm{CRB}_{\mathrm{r}}(\bm{q}).
\end{equation}
The inequality is strict whenever the \ac{DT} fingerprint exhibits nonzero spatial variation, i.e., $\partial \bm{f}_{\mathrm{d}}(\bm{q})/\partial \bm{q} \neq \bm{0}$.
The total \ac{DT} information $\tr(\bm{J}_{\mathrm{d}}(\bm{q})) = {\left\|\partial \bm{f}_{\mathrm{d}}(\bm{q})/\partial \bm{q}\right\|_F^2}/{\sigma_{\Delta}^2}$, the \ac{DT}'s spatial information relative to its domain gap.
\end{proposition}

\begin{proof}
The proof proceeds in three steps.

\emph{Step~1 (Positive semidefiniteness).}
By~\eqref{eq:fim_dt}, $\bm{x}^{\top}\bm{J}_{\mathrm{d}}(\bm{q})\,\bm{x} = \|(\partial \bm{f}_{\mathrm{d}}(\bm{q})/\partial \bm{q})\,\bm{x}\|^2/\sigma_{\Delta}^2 \geq 0,\ \forall \bm{x} \in \real^2$, so $\bm{J}_{\mathrm{d}}(\bm{q}) \succeq \bm{0}$.

\emph{Step~2 (Non-increasing CRB).}
With Step~1 and $\bm{J}_{\mathrm{r}}(\bm{q}) \succ \bm{0}$, from~\eqref{eq:fim_joint},
\begin{equation}
\bm{J}_{\mathrm{r\!+\!d}}(\bm{q}) = \bm{J}_{\mathrm{r}}(\bm{q}) + \bm{J}_{\mathrm{d}}(\bm{q}) \succeq \bm{J}_{\mathrm{r}}(\bm{q}) \succ \bm{0}.
\end{equation}
By the L\"{o}wner ordering ($\bm{A} \succeq \bm{B} \succ \bm{0} \Rightarrow \bm{A}^{-1} \preceq \bm{B}^{-1}$),
\begin{equation}
\bm{J}_{\mathrm{r\!+\!d}}^{-1}(\bm{q}) \preceq \bm{J}_{\mathrm{r}}^{-1}(\bm{q})
\;\Longrightarrow\;
\tr\!\left(\bm{J}_{\mathrm{r\!+\!d}}^{-1}(\bm{q})\right) \leq \tr\!\left(\bm{J}_{\mathrm{r}}^{-1}(\bm{q})\right),
\end{equation}
which gives $\mathrm{CRB}_{\mathrm{r\!+\!d}}(\bm{q}) \leq \mathrm{CRB}_{\mathrm{r}}(\bm{q})$.

\emph{Step~3 (Strict improvement condition).}
With $\bm{\Gamma} = \bm{J}_{\mathrm{r}}^{-1/2}\bm{J}_{\mathrm{d}}\bm{J}_{\mathrm{r}}^{-1/2} \succeq \bm{0}$ (argument $\bm{q}$ omitted), \eqref{eq:fim_joint} gives \begin{equation*} \bm{J}_{\mathrm{r}}^{-1} - \bm{J}_{\mathrm{r\!+\!d}}^{-1} = \bm{J}_{\mathrm{r}}^{-1/2}\bm{\Gamma}(\bm{I}_2 + \bm{\Gamma})^{-1}\bm{J}_{\mathrm{r}}^{-1/2} \succeq \bm{0}, \end{equation*} so $\tr(\bm{J}_{\mathrm{r}}^{-1}) > \tr(\bm{J}_{\mathrm{r\!+\!d}}^{-1}) \Leftrightarrow \bm{\Gamma} \neq \bm{0} \Leftrightarrow \partial \bm{f}_{\mathrm{d}}(\bm{q})/\partial \bm{q} \neq \bm{0}$.
Since $\tr(\bm{A}^{\top}\bm{A}) = \|\bm{A}\|_F^2$, \eqref{eq:fim_dt} gives $\tr(\bm{J}_{\mathrm{d}}(\bm{q})) = {\left\|\partial \bm{f}_{\mathrm{d}}(\bm{q})/\partial \bm{q}\right\|_F^2}/{\sigma_{\Delta}^2}$.
\end{proof}

\begin{remark}[Directional complementarity]
\label{rem:crb_directional}
To characterize \emph{where} the \ac{DT} information is most valuable, we analyze the first-order sensitivity of the \ac{CRB} to the \ac{DT} contribution. Define
\begin{equation}
\label{eq:sensitivity}
\mathcal{S}(\beta,\bm{q}) = \tr\!\left(\left(\bm{J}_{\mathrm{r}}(\bm{q}) + \beta \bm{J}_{\mathrm{d}}(\bm{q})\right)^{-1}\right), \quad \beta \geq 0.
\end{equation}
Then
\begin{equation}
\label{eq:sensitivity_deriv}
\left.\frac{\partial \mathcal{S}}{\partial \beta}\right|_{\beta=0} = -\tr\!\left(\bm{J}_{\mathrm{r}}^{-1}(\bm{q})\,\bm{J}_{\mathrm{d}}(\bm{q})\,\bm{J}_{\mathrm{r}}^{-1}(\bm{q})\right) = -\sum_{i=1}^{2} \frac{\bm{c}_i^{\top}\bm{J}_{\mathrm{d}}(\bm{q})\,\bm{c}_i}{\lambda_i^2},
\end{equation}
by $\partial \bm{X}^{-1}/\partial \beta = -\bm{X}^{-1}(\partial \bm{X}/\partial \beta)\bm{X}^{-1}$, where $(\lambda_i,\bm{c}_i)$, $\lambda_1 \geq \lambda_2 > 0$, are the eigenpairs of $\bm{J}_{\mathrm{r}}(\bm{q})$. Hence, the largest first-order \ac{CRB} reduction occurs when the \ac{DT} adds information along the weak eigendirection $\bm{c}_2$ of the real-only \ac{FIM}. On our testbed, 73\% of the \ac{DT}'s \ac{CRB} reduction comes from the weak eigendirection, so the \ac{DT} acts mainly along $\bm{c}_2$, motivating position-dependent \ac{AP} weighting.
\end{remark}

\begin{corollary}[Per-\acs{AP} marginal \acs{DT} utility]
\label{cor:per_ap}
Since $\bm{f}_{\mathrm{d}} = [\bm{f}_{\mathrm{d}}^{(1)\top}, \ldots, \bm{f}_{\mathrm{d}}^{(M)\top}]^{\top}$ with $\bm{f}_{\mathrm{d}}^{(m)} = \bm{d}^{(m)}$, the \ac{DT} \ac{FIM}~\eqref{eq:fim_dt} decomposes as $\bm{J}_{\mathrm{d}}(\bm{q}) = \sum_{m=1}^{M} \bm{J}_{\mathrm{d}}^{(m)}(\bm{q})$, where
\begin{equation}
\label{eq:fim_dt_ap}
\bm{J}_{\mathrm{d}}^{(m)}(\bm{q}) = \frac{1}{\sigma_{\Delta}^2} \left(\frac{\partial \bm{f}_{\mathrm{d}}^{(m)}(\bm{q})}{\partial \bm{q}}\right)^{\!\!\top} \frac{\partial \bm{f}_{\mathrm{d}}^{(m)}(\bm{q})}{\partial \bm{q}}.
\end{equation}
The first-order \ac{CRB} sensitivity from Remark~\ref{rem:crb_directional} decomposes accordingly as $-\partial \mathcal{S}/\partial\beta|_{\beta=0} = \sum_{m=1}^{M} \mu_m(\bm{q})$, where the \emph{per-\acs{AP} marginal \acs{DT} utility} is
\begin{equation}
\label{eq:per_ap_utility}
\mu_m(\bm{q}) = \tr\!\left(\bm{J}_{\mathrm{r}}^{-1}(\bm{q})\, \bm{J}_{\mathrm{d}}^{(m)}(\bm{q})\, \bm{J}_{\mathrm{r}}^{-1}(\bm{q})\right) = \sum_{i=1}^{2} \frac{\bm{c}_i^{\top} \bm{J}_{\mathrm{d}}^{(m)}(\bm{q})\, \bm{c}_i}{\lambda_i^2}.
\end{equation}
\looseness=-1 The score $\mu_m(\bm{q})$ quantifies how much \ac{AP}~$m$'s \ac{DT} fingerprint reduces the oracle \ac{CRB} at position~$\bm{q}$, weighted by the real-only information geometry. In cell-free deployments, the geometric diversity of \ac{AP} placements ensures that per-\acs{AP} \ac{DT} Jacobians point in complementary directions in position space, so even an \ac{AP} with low absolute \ac{DT} fidelity ($\rho_m$) can contribute substantially if its \ac{DT} Jacobian complements the information provided by other \acp{AP}. The learned \ac{AP} attention weights $\alpha_m$ in \acs{DTCond} (Section~\ref{sec:framework_attention}) are designed as a data-driven proxy for $\mu_m(\bm{q})$, which Section~\ref{sec:exp_crb} evaluates.
\end{corollary}

\begin{remark}[Cell-free geometric diversity gain]
\label{rem:cellfree_diversity}
In a cell-free deployment, the per-\acs{AP} \ac{DT} Jacobians $\partial \bm{f}_{\mathrm{d}}^{(m)}(\bm{q})/\partial \bm{q}$ point in geometrically diverse directions because the \acp{AP} observe the \ac{UE} from different angles, ensuring that $\bm{J}_{\mathrm{d}}(\bm{q}) = \sum_{m=1}^{M}\bm{J}_{\mathrm{d}}^{(m)}(\bm{q}) \succ \bm{0}$, so $\bm{c}_2^{\top}\bm{J}_{\mathrm{d}}(\bm{q})\,\bm{c}_2 > 0$ along the weak eigendirection. The \ac{DT} thus improves both the size and shape of the positioning error ellipse, a benefit that grows with the angular diversity of distributed \acp{AP}.
\end{remark}

\begin{remark}[Quantifying the improvement]
\label{rem:crb_quantify}
When the real and \ac{DT} fingerprint Jacobians are aligned ($\partial \bm{f}_{\mathrm{d}}(\bm{q})/\partial \bm{q} = c\, \partial \bm{f}_{\mathrm{r}}(\bm{q})/\partial \bm{q}$ for some $c > 0$), then $\bm{J}_{\mathrm{d}} = (c^2\sigma_{\mathrm{r}}^2/\sigma_{\Delta}^2)\bm{J}_{\mathrm{r}}$ by~\eqref{eq:fim_real} and~\eqref{eq:fim_dt}, and the joint \ac{CRB} simplifies to
\begin{equation}
\label{eq:crb_aligned}
\mathrm{CRB}_{\mathrm{r\!+\!d}}(\bm{q}) = \frac{\mathrm{CRB}_{\mathrm{r}}(\bm{q})}{\sqrt{1 + c^2\, \sigma_{\mathrm{r}}^2 / \sigma_{\Delta}^2}}.
\end{equation}

This shows that the improvement depends on both the Jacobian alignment factor $c$ and the noise-to-gap ratio $\sigma_{\mathrm{r}}^2 / \sigma_{\Delta}^2$.
More generally, the \ac{DT} is most valuable when measurement noise is large relative to the domain gap, which is precisely the regime where real-only localization struggles.
In practice, the alignment factor is estimated from finite data as
\begin{equation}
\label{eq:c_hat}
\hat{c}(\bm{q}_k) = \operatorname*{arg\,min}_{c} \|\bm{F}_{\mathrm{d}}^{(k)} - c\,\bm{F}_{\mathrm{r}}^{(k)}\|_F^2 = \frac{\langle \bm{F}_{\mathrm{r}}^{(k)},\, \bm{F}_{\mathrm{d}}^{(k)} \rangle_F}{\|\bm{F}_{\mathrm{r}}^{(k)}\|_F^2},
\end{equation}
where $\bm{F}_{\mathrm{r}}^{(k)}$ and $\bm{F}_{\mathrm{d}}^{(k)}$ are the real and \ac{DT} fingerprint Jacobians evaluated at position~$\bm{q}_k$.
Section~\ref{sec:exp_crb} reports the empirical distribution of $\hat{c}$ and validates this prediction against the measured \ac{CRB} reduction.

\end{remark}

\begin{proposition}[Heteroscedastic per-AP CRB]
\label{prop:het_crb}
Let $\sigma^2_{\Delta,m}(\bm{q})$ denote the domain-gap variance, treated as locally constant ($\partial \sigma^2_{\Delta,m}(\bm{q})/\partial \bm{q} = \bm{0}$), and define the heteroscedastic DT FIM
\begin{equation}
\bm{J}_{\mathrm{d,het}}(\bm{q}) = \sum_{m=1}^{M} \frac{1}{\sigma^2_{\Delta,m}(\bm{q})} \bm{G}_m^\top \bm{G}_m, \quad \bm{G}_m = \frac{\partial \bm{f}^{(m)}_{\mathrm{d}}(\bm{q})}{\partial \bm{q}}.
\end{equation}
If $\bm{J}_{\mathrm{r}}(\bm{q}) \succ \bm{0}$ and $\bm{J}_{\mathrm{d,het}}(\bm{q}) \succeq \bm{J}_{\mathrm{d}}(\bm{q})$, then the heteroscedastic CRB, i.e.,~\eqref{eq:crb} with $\bm{J}_{\mathrm{d,het}}$ in place of $\bm{J}_{\mathrm{d}}$, satisfies
\begin{equation}
\mathrm{CRB}_{\mathrm{het}}(\bm{q}) \leq \mathrm{CRB}_{\mathrm{r\!+\!d}}(\bm{q}) \leq \mathrm{CRB}_{\mathrm{r}}(\bm{q}).
\end{equation}
A sufficient condition for $\bm{J}_{\mathrm{d,het}}(\bm{q}) \succeq \bm{J}_{\mathrm{d}}(\bm{q})$ is $\sigma^2_{\Delta,m}(\bm{q}) \leq \sigma^2_\Delta,\ \forall m$; if $\sigma^2_{\Delta,m}(\bm{q}) = \sigma^2_\Delta,\ \forall m$, then $\mathrm{CRB}_{\mathrm{het}}(\bm{q}) = \mathrm{CRB}_{\mathrm{r\!+\!d}}(\bm{q})$.
\end{proposition}

\begin{proof}
When $\bm{J}_{\mathrm{d,het}}(\bm{q}) \succeq \bm{J}_{\mathrm{d}}(\bm{q})$, $\bm{J}_{\mathrm{r}} + \bm{J}_{\mathrm{d,het}} \succeq \bm{J}_{\mathrm{r}} + \bm{J}_{\mathrm{d}} \succeq \bm{J}_{\mathrm{r}} \succ \bm{0}$, and the L\"owner ordering on positive-definite inverses yields $\tr((\bm{J}_{\mathrm{r}} + \bm{J}_{\mathrm{d,het}})^{-1}) \leq \tr((\bm{J}_{\mathrm{r}} + \bm{J}_{\mathrm{d}})^{-1})$. The sufficient condition follows because $\sigma^2_{\Delta,m}(\bm{q}) \leq \sigma^2_\Delta,\ \forall m$, implies $\bm{J}_{\mathrm{d,het}} - \bm{J}_{\mathrm{d}} = \sum_{m=1}^{M} (\sigma^{-2}_{\Delta,m}(\bm{q}) - \sigma^{-2}_{\Delta})\,\bm{G}_m^{\top}\bm{G}_m \succeq \bm{0}$, using $\bm{J}_{\mathrm{d}} = \sum_{m=1}^{M} \sigma_{\Delta}^{-2}\bm{G}_m^{\top}\bm{G}_m$ (Corollary~\ref{cor:per_ap}) and $\bm{G}_m^{\top}\bm{G}_m \succeq \bm{0}$. Local constancy omits the covariance-derivative term $\bm{J}_{\Sigma}(\bm{q}) \succeq \bm{0}$ of the Gaussian FIM; since $\bm{J}_{\mathrm{d,het}} + \bm{J}_{\Sigma} \succeq \bm{J}_{\mathrm{d,het}} \succeq \bm{J}_{\mathrm{d}}$, the ordering persists. The right inequality is Proposition~\ref{prop:crb}.
\end{proof}

\looseness=-1 On our testbed, the heteroscedastic bound is tighter than the uniform bound at 97.1\% of test positions (Section~\ref{sec:exp_crb}), and the mean heteroscedastic CRB (0.162~m) is 9.1\% tighter than the uniform bound (0.178~m). Proposition~\ref{prop:het_crb} connects this refinement to the learnable AP attention in Section~\ref{sec:framework_attention}: an architecture with per-AP precision weights can approach the heteroscedastic bound at positions where the condition holds.

\begin{proposition}[Pipeline \acs{CRB} under \acs{DT} retrieval mismatch]
\label{prop:pipeline_crb}
At deployment, the \ac{DT} fingerprint is retrieved at a coarse estimate $\hat{\bm{q}}_0$, introducing a retrieval error $\Delta\bm{q} = \hat{\bm{q}}_0 - \bm{q} \sim \mathcal{N}(\bm{0}, \bm{C}_0)$, independent of the noise $\bm{n}$ and the domain gap $\bm{\delta}(\bm{q})$, with $\bm{G}$ and $\bm{C}_0$ locally constant as in Proposition~\ref{prop:het_crb}. Under a first-order Taylor approximation $\bm{f}_{\mathrm{d}}(\hat{\bm{q}}_0) \approx \bm{f}_{\mathrm{d}}(\bm{q}) + (\partial \bm{f}_{\mathrm{d}}(\bm{q})/\partial \bm{q})\,\Delta\bm{q}$, the retrieval mismatch inflates the effective \ac{DT} observation noise to
\begin{equation}
\label{eq:sigma_eff}
\bm{\Sigma}_{\mathrm{eff}}(\bm{q}) = \sigma_{\Delta}^2 \bm{I} + \frac{\partial \bm{f}_{\mathrm{d}}(\bm{q})}{\partial \bm{q}}\, \bm{C}_0 \left(\frac{\partial \bm{f}_{\mathrm{d}}(\bm{q})}{\partial \bm{q}}\right)^{\!\!\top}\!,
\end{equation}
yielding the retrieved-\acs{DT} \ac{FIM} contribution
\begin{equation}
\label{eq:fim_dt_ret}
\bm{J}_{\mathrm{d,ret}}(\bm{q}) = \left(\frac{\partial \bm{f}_{\mathrm{d}}(\bm{q})}{\partial \bm{q}}\right)^{\!\!\top} \bm{\Sigma}_{\mathrm{eff}}^{-1}(\bm{q})\, \frac{\partial \bm{f}_{\mathrm{d}}(\bm{q})}{\partial \bm{q}}.
\end{equation}
The pipeline CRB is 
\begin{equation} \label{eq:crb_pipe_def} \mathrm{CRB}_{\mathrm{pipe}}(\bm{q};\bm{C}_0) = \sqrt{\tr\!\left((\bm{J}_{\mathrm{r}}(\bm{q}) + \bm{J}_{\mathrm{d,ret}}(\bm{q}))^{-1}\right)}, \end{equation} 
which satisfies the sandwich inequality
\begin{equation}
\label{eq:sandwich}
\mathrm{CRB}_{\mathrm{r\!+\!d}}(\bm{q}) \leq \mathrm{CRB}_{\mathrm{pipe}}(\bm{q};\bm{C}_0) \leq \mathrm{CRB}_{\mathrm{r}}(\bm{q}).
\end{equation}
When the coarse-stage error is the real-only \ac{CRB} ($\bm{C}_0 = \bm{J}_{\mathrm{r}}^{-1}(\bm{q})$), the degradation from oracle to pipeline is governed by the ratio $\|\bm{G}\bm{J}_{\mathrm{r}}^{-1}(\bm{q})\bm{G}^{\top}\|_2 / \sigma_{\Delta}^2$, where $\bm{G} = \partial \bm{f}_{\mathrm{d}}(\bm{q})/\partial \bm{q} = [\bm{G}_1^{\top},\, \ldots,\, \bm{G}_M^{\top}]^{\top} \in \mathbb{R}^{MN_{\mathrm{b}} \times 2}$ stacks the per-\ac{AP} Jacobians from Proposition~\ref{prop:het_crb}. The pipeline approaches the oracle when the coarse stage is accurate or the \ac{DT} fingerprint varies slowly, and approaches the real-only bound when the retrieval mismatch dominates the domain gap.
Moreover, with $\bm{J}_{\mathrm{r}}(\bm{q}) \succ \bm{0}$,
(a)~$\bm{J}_{\mathrm{d,ret}} = \bm{J}_{\mathrm{d}}(\bm{I}_2 + \bm{C}_0\bm{J}_{\mathrm{d}})^{-1}$,
(b)~$\mathrm{CRB}_{\mathrm{pipe}}(\bm{q};\bm{C}_0) \leq \mathrm{CRB}_{\mathrm{pipe}}(\bm{q};\bm{C}_0'),\ \forall \bm{C}_0' \succeq \bm{C}_0$, and
(c)~$\mathrm{CRB}_{\mathrm{pipe}}(\bm{q};\bm{0}) = \mathrm{CRB}_{\mathrm{r\!+\!d}}(\bm{q})$, whereas $\mathrm{CRB}_{\mathrm{pipe}}(\bm{q};\tau\bm{C}) \to \mathrm{CRB}_{\mathrm{r}}(\bm{q})$ as $\tau \to \infty$, $\forall \bm{C} \succ \bm{0}$.
\end{proposition}

\begin{proof}
The effective noise covariance satisfies $\bm{\Sigma}_{\mathrm{eff}}(\bm{q}) = \sigma_{\Delta}^2 \bm{I} + \bm{G}\bm{C}_0\bm{G}^{\top} \succeq \sigma_{\Delta}^2 \bm{I}$ because $\bm{C}_0 \succeq \bm{0}$. By the L\"{o}wner ordering on inverses, $\bm{\Sigma}_{\mathrm{eff}}^{-1}(\bm{q}) \preceq \sigma_{\Delta}^{-2}\bm{I}$, which gives $\bm{J}_{\mathrm{d,ret}}(\bm{q}) = \bm{G}^{\top}\bm{\Sigma}_{\mathrm{eff}}^{-1}(\bm{q})\,\bm{G} \preceq \bm{G}^{\top}(\sigma_{\Delta}^{-2}\bm{I})\bm{G} = \bm{J}_{\mathrm{d}}(\bm{q})$. Hence $\bm{J}_{\mathrm{r}} + \bm{J}_{\mathrm{d,ret}} \preceq \bm{J}_{\mathrm{r}} + \bm{J}_{\mathrm{d}}$, and Step~2 of the proof of Proposition~\ref{prop:crb} yields the left inequality. The right inequality follows because $\bm{J}_{\mathrm{d,ret}}(\bm{q}) \succeq \bm{0}$ (a congruence of $\bm{\Sigma}_{\mathrm{eff}}^{-1}(\bm{q}) \succ \bm{0}$), so $\bm{J}_{\mathrm{r}}(\bm{q}) + \bm{J}_{\mathrm{d,ret}}(\bm{q}) \succeq \bm{J}_{\mathrm{r}}(\bm{q})$.
(a)~With $\bm{G}^{\top}\bm{G} = \sigma_{\Delta}^2\bm{J}_{\mathrm{d}}$, the push-through identity $(\bm{I} + \bm{S}\bm{T})^{-1}\bm{S} = \bm{S}(\bm{I} + \bm{T}\bm{S})^{-1}$ with $\bm{S} = \bm{G}^{\top}$ and $\bm{T} = \sigma_{\Delta}^{-2}\bm{G}\bm{C}_0$ gives $\bm{J}_{\mathrm{d,ret}} = (\bm{I}_2 + \bm{J}_{\mathrm{d}}\bm{C}_0)^{-1}\bm{J}_{\mathrm{d}} = \bm{J}_{\mathrm{d}}(\bm{I}_2 + \bm{C}_0\bm{J}_{\mathrm{d}})^{-1}$.
(b)~$\bm{C}_0 \preceq \bm{C}_0'$ gives $\bm{\Sigma}_{\mathrm{eff}} \preceq \bm{\Sigma}_{\mathrm{eff}}'$ in~\eqref{eq:sigma_eff}, hence $\bm{J}_{\mathrm{d,ret}} \succeq \bm{J}_{\mathrm{d,ret}}'$, and the ordering follows as in Proposition~\ref{prop:crb}.
(c)~$\bm{C}_0 = \bm{0}$ gives $\bm{J}_{\mathrm{d,ret}} = \bm{J}_{\mathrm{d}}$. For $\bm{C}_0 = \tau\bm{C} \succeq \tau\lambda_{\min}(\bm{C})\bm{I}_2$, parts (b) and (a) give $\bm{0} \preceq \bm{J}_{\mathrm{d,ret}} \preceq \bm{J}_{\mathrm{d}}(\bm{I}_2 + \tau\lambda_{\min}(\bm{C})\bm{J}_{\mathrm{d}})^{-1} \to \bm{0}$.
\end{proof}

Proposition~\ref{prop:pipeline_crb} yields three insights. First, the sandwich inequality~\eqref{eq:sandwich} places the pipeline bound between the oracle \ac{CRB} and the real-only bound, so the \ac{DT} never loosens the bound, even when read at a coarse estimate; Section~\ref{sec:exp_crb} evaluates it at all test positions. Second, iterative refinement (re-retrieving the \ac{DT} at the \acs{DTCond} output) diverges because it feeds dependent estimation errors back through $\bm{G}$, whereas the single-pass pipeline reads the \ac{DT} once, at a real-only \ac{kNN} coarse estimate. Third, the ratio $\|\bm{G}\bm{J}_{\mathrm{r}}^{-1}(\bm{q})\bm{G}^{\top}\|_2/\sigma_{\Delta}^2$ is a deployable diagnostic: when small, the pipeline approaches oracle performance; when large, improving the coarse stage yields higher returns than improving the \ac{DT}.

\looseness=-1 A sparser site survey (the real fingerprint database of the coarse \ac{kNN} stage) enlarges $\bm{C}_0$ and, by part~(b), does not lower the pipeline bound, since nearest-node retrieval on a lattice with spacings $h_x$ and $h_y$ contributes $\diag(h_x^2, h_y^2)/12$.

The standard \ac{CRB} assumes Gaussian observation noise.
We verify empirically in Section~\ref{sec:exp_crb} that mild departures from Gaussianity do not materially affect the bound, using the misspecified \ac{CRB} framework of~\cite{richmond2015misspecified}.

\section{\acs{DT}-Conditioned Localization Framework}
\label{sec:framework}

\begin{figure*}[!t]
\centering
\includegraphics[width=\textwidth]{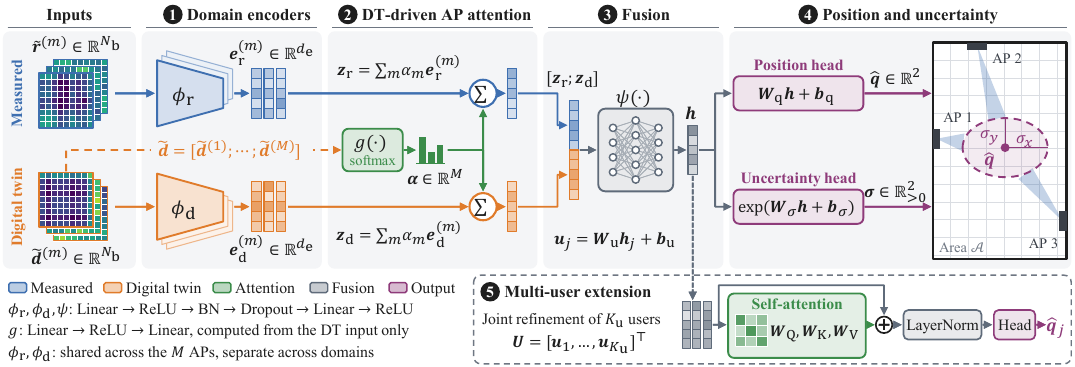}
\caption{DTCond architecture with per-domain encoders, \ac{DT}-driven \ac{AP} attention, and heteroscedastic position output.}
\label{fig:architecture}
\end{figure*}

Motivated by the \ac{CRB} analysis, which shows that the \ac{DT} contributes positioning information through spatial structure rather than absolute accuracy, we design a localization architecture that \emph{conditions} on the \ac{DT} input while preserving the distinct statistical characteristics of each domain.
The architecture, illustrated in Fig.~\ref{fig:architecture}, processes real and \ac{DT} fingerprints through separate encoder pathways, fuses them via learnable \ac{AP} attention, and outputs a position estimate with learned uncertainty.

\vspace{-2mm}
\subsection{Data Preprocessing Pipeline}
\label{sec:system_preprocessing}

Raw linear power measurements are first converted to dBm via $r^{(m)}_{n,\mathrm{dBm}} = 10 \log_{10}(\max(P_{m,n}, \epsilon))$, where $\epsilon > 0$ prevents logarithmic singularities.
Per-\ac{AP} offsets $\Delta P_m$ (in~dB), added to both domains, balance the \acp{AP}: with $N_m = |\{k : m = \argmax_{m'} (\max_{n} r^{(m')}_{n}(\bm{q}_k) + \Delta P_{m'})\}|$ the number of positions at which \ac{AP}~$m$ provides the strongest beam pair, the offsets satisfy $\max_m |N_m/N_{\mathrm{p}} - 1/M| < 0.02$, giving 32.3, 34.9, and 32.8\% for \acp{AP}~1--3 instead of 6.2, 88.0, and 5.7\%.
This prevents a proximate \ac{AP} from dominating and masking the spatial information carried by more distant \acp{AP}.
The offsets and the normalization statistics use the fingerprints without position labels, and offsets computed on the training split alone differ by less than 0.05\,dB.
Finally, each domain is z-score normalized using its own per-\ac{AP} statistics
\begin{equation}
\label{eq:normalize}
\tilde{\bm{r}}^{(m)} = \frac{\bm{r}^{(m)} - \bar{r}^{(m)}\bm{1}}{\sigma_{\mathrm{r,stat}}^{(m)}}, \quad
\tilde{\bm{d}}^{(m)} = \frac{\bm{d}^{(m)} - \bar{d}^{(m)}\bm{1}}{\sigma_{\mathrm{d,stat}}^{(m)}},
\end{equation}
where $\bar{r}^{(m)}$ and $\sigma_{\mathrm{r,stat}}^{(m)}$ are the mean and standard deviation of the entries of $\bm{r}^{(m)}$ over all beam pairs and positions, and the \ac{DT} statistics are obtained in the same way.
Per-domain normalization is essential; sharing statistics across domains degrades accuracy because the \ac{DT}'s different dynamic range distorts the real signal statistics.

\subsection{Per-Domain Encoders}
\label{sec:framework_encoders}

For each \ac{AP}~$m \in \{1, \ldots, M\}$, separate encoders transform the per-\ac{AP} normalized fingerprints into compact representations. Let $\bm{e}_{\mathrm{r}}^{(m)} \in \real^{d_{\mathrm{e}}}$ and $\bm{e}_{\mathrm{d}}^{(m)} \in \real^{d_{\mathrm{e}}}$ denote the real-domain and \ac{DT}-domain embedding vectors for \ac{AP}~$m$, respectively, where $d_{\mathrm{e}}$ is the embedding dimension. These are obtained by applying per-domain encoder networks $\phi_{\mathrm{r}}$ and $\phi_{\mathrm{d}}$ to the corresponding normalized fingerprints
\begin{equation}
\label{eq:encoders}
\bm{e}_{\mathrm{r}}^{(m)} = \phi_{\mathrm{r}}(\tilde{\bm{r}}^{(m)}), \quad \bm{e}_{\mathrm{d}}^{(m)} = \phi_{\mathrm{d}}(\tilde{\bm{d}}^{(m)}).
\end{equation}
Both encoders $\phi_{\mathrm{r}}$ and $\phi_{\mathrm{d}}$ share the same two-layer \ac{MLP} architecture, written generically as $\phi_i$, $i \in \{\mathrm{r},\, \mathrm{d}\}$:
\begin{equation}
\label{eq:encoder_arch}
\begin{aligned}
\phi_i(\bm{x}) = \mathrm{ReLU}\!\Bigl(
&\bm{W}_2\,\mathrm{Dropout}_{p_{\mathrm{d}}}\!\bigl(
\mathrm{BN}\!\bigl(\mathrm{ReLU}(\bm{W}_1 \bm{x} + \bm{b}_1)\bigr)
\bigr) \\
&+ \bm{b}_2
\Bigr),
\end{aligned}
\end{equation}
where $\bm{W}_1 \in \real^{d_{\mathrm{h}} \times N_{\mathrm{b}}}$, $\bm{W}_2 \in \real^{d_{\mathrm{e}} \times d_{\mathrm{h}}}$, $d_{\mathrm{h}}$ is the hidden layer dimension, $\mathrm{BN}$ denotes \ac{BN}, and $\mathrm{Dropout}_{p_{\mathrm{d}}}$ applies dropout with probability~$p_{\mathrm{d}}$.
The encoders are \emph{shared across \acp{AP}} (i.e., the same $\phi_{\mathrm{r}}$ processes $\tilde{\bm{r}}^{(1)}, \ldots, \tilde{\bm{r}}^{(M)}$) but \emph{separate across domains} ($\phi_{\mathrm{r}} \neq \phi_{\mathrm{d}}$).
\looseness=-1 Weight sharing across \acp{AP} reduces parameters from $\mathcal{O}(M)$ to $\mathcal{O}(1)$ encoder sets, while domain separation allows each encoder to learn domain-appropriate feature extraction, respecting the different distributional characteristics of real and \ac{DT} fingerprints.

\subsection{Learnable \acs{AP} Attention}
\label{sec:framework_attention}

Not all \acp{AP} contribute equally to localization at every position.
An \ac{AP} with a strong \ac{LoS} path to the \ac{UE} provides more discriminative beam-domain information than one whose signal is heavily shadowed or arrives through multiple reflections.
We compute position-dependent \ac{AP} attention weights $\bm{\alpha} = [\alpha_1,\, \dots,\, \alpha_M]^{\top} \in \real^M$, where $\alpha_m$ denotes the scalar weight assigned to \ac{AP}~$m$, from the \ac{DT} input
\begin{equation}
\label{eq:attention}
\bm{\alpha} = \softmax\!\left(g(\tilde{\bm{d}})\right) \in \real^M, \quad \alpha_m \geq 0 \; \forall m, \; \sum_{m=1}^{M} \alpha_m = 1,
\end{equation}
where $g: \real^{MN_{\mathrm{b}}} \to \real^M$ is a two-layer network, $\mathrm{Linear}(MN_{\mathrm{b}} \to d_{\mathrm{a}}) \to \mathrm{ReLU} \to \mathrm{Linear}(d_{\mathrm{a}} \to M)$, with hidden dimension $d_{\mathrm{a}}$.
Computing $\bm{\alpha}$ exclusively from the \ac{DT} fingerprint enables attention computation where no real measurements exist (useful during progressive calibration) and lets the network learn which \acp{AP} are most informative at each region from the \ac{DT}'s beam patterns, serving as a data-driven proxy for the per-\ac{AP} \ac{DT} utility of Corollary~\ref{cor:per_ap}.

The weighted aggregation produces the real-domain global feature vector $\bm{z}_{\mathrm{r}} \in \real^{d_{\mathrm{e}}}$ and the \ac{DT}-domain global feature vector $\bm{z}_{\mathrm{d}} \in \real^{d_{\mathrm{e}}}$, computed as
\begin{equation}
\label{eq:aggregation}
\bm{z}_{\mathrm{r}} = \sum_{m=1}^{M} \alpha_m \, \bm{e}_{\mathrm{r}}^{(m)}, \quad
\bm{z}_{\mathrm{d}} = \sum_{m=1}^{M} \alpha_m \, \bm{e}_{\mathrm{d}}^{(m)}.
\end{equation}

\subsection{Fusion and Heteroscedastic Uncertainty}
\label{sec:framework_fusion}

The fusion network combines the domain-specific features and jointly predicts position and uncertainty
\begin{equation}
\label{eq:fusion}
\bm{h} = \psi\!\left([\bm{z}_{\mathrm{r}};\, \bm{z}_{\mathrm{d}}]\right) \in \real^{d_{\mathrm{f}}},
\end{equation}
where $\psi: \real^{2d_{\mathrm{e}}} \to \real^{d_{\mathrm{f}}}$ is a two-layer \ac{MLP} with the same structure as~\eqref{eq:encoder_arch} with input dimension $2d_{\mathrm{e}}$, hidden dimension $d_{\mathrm{h}}$, and output dimension $d_{\mathrm{f}}$.
A linear output head maps from the fused representation to position and log-uncertainty
\begin{equation}
\label{eq:output_head}
\hat{\bm{q}} = \bm{W}_{\mathrm{q}} \bm{h} + \bm{b}_{\mathrm{q}} \in \real^2, \quad
\log\bm{\sigma} = \bm{W}_{\sigma} \bm{h} + \bm{b}_{\sigma} \in \real^2,
\end{equation}
where $\bm{W}_{\mathrm{q}} \in \real^{2 \times d_{\mathrm{f}}}$ and $\bm{b}_{\mathrm{q}} \in \real^2$ are the weight matrix and bias vector of the position head, and $\bm{W}_{\sigma} \in \real^{2 \times d_{\mathrm{f}}}$ and $\bm{b}_{\sigma} \in \real^2$ are the corresponding parameters of the uncertainty head. Denoting the network output as $\bm{s} = \bm{W}_{\sigma} \bm{h} + \bm{b}_{\sigma} \in \real^2$, the per-coordinate predicted uncertainty is obtained as $\bm{\sigma} = [\sigma_x,\, \sigma_y]^{\top} = \exp(\bm{s}) \in \real_{>0}^2$, where the elementwise exponential ensures positivity.
The heteroscedastic output $\bm{\sigma}(\bm{q})$ provides a spatially varying confidence measure, with small values in well-covered regions and large values in undersampled or ambiguous areas.

\subsection{Two-Stage Training}
\label{sec:framework_training}

Training proceeds in two stages to decouple position learning from uncertainty calibration.

\textbf{Stage~1, \acs{MSE} warmup ($T_{\mathrm{w}}$ epochs).}
The network learns positions under the \ac{MSE} loss, ignoring the uncertainty outputs

\begin{equation}
\label{eq:mse_loss}
\mathcal{L}_{\mathrm{MSE}} = \frac{1}{N_{\mathrm{c}}} \sum_{k=1}^{N_{\mathrm{c}}} \left\|\hat{\bm{q}}_k - \bm{q}_k\right\|^2.
\end{equation}
This stage establishes stable position estimates before introducing the more complex uncertainty objective.
The Adam optimizer is used with gradient clipping at $\|\nabla_{\bm{\theta}}\mathcal{L}\| \leq \nabla_{\max}$ to prevent instability from noisy gradients; training hyperparameters are listed with the system parameters in Section~\ref{sec:exp_setup}.

\textbf{Stage~2, heteroscedastic \acs{NLL} ($T_{\mathrm{n}}$ epochs).}
The loss transitions to the heteroscedastic \ac{NLL}~\cite{kendall2017uncertainty}

\begin{equation}
\label{eq:nll_loss}
\mathcal{L}_{\mathrm{NLL}} = \frac{1}{N_{\mathrm{c}}} \sum_{k=1}^{N_{\mathrm{c}}} \sum_{i \in \{x,y\}} \left[\log \sigma_{k,i} + \frac{(\hat{q}_{k,i} - q_{k,i})^2}{2\sigma_{k,i}^2}\right],
\end{equation}
which jointly refines position accuracy and calibrates the uncertainty output. 
The quadratic term penalizes errors that are large relative to $\bm{\sigma}$, i.e., overconfident predictions, and the logarithmic term penalizes needlessly large $\bm{\sigma}$.
Cosine annealing reduces the learning rate over $T_{\mathrm{w}} + T_{\mathrm{n}}$ total epochs, with early stopping on the validation \ac{MSE}.

The two-stage approach prevents \emph{uncertainty collapse}, where the model minimizes \ac{NLL} by predicting uniformly large $\sigma$ everywhere rather than learning accurate positions.
By first establishing good position estimates (Stage~1), the uncertainty head in Stage~2 can learn spatially varying confidence that correlates with actual error.

\subsection{Multi-User Localization via Self-Attention}
\label{sec:framework_multiuser}

When $K_{\mathrm{u}} > 1$ users are simultaneously present in $\mathcal{A}$, cross-user information can improve per-user accuracy~\cite{win2011cooperative}.
Each user $j \in \{1, \ldots, K_{\mathrm{u}}\}$ independently passes through the shared \acs{DTCond} pipeline (encoders, attention, fusion) to obtain a per-user fusion output $\bm{h}_j = \psi([\bm{z}_{\mathrm{r},j};\, \bm{z}_{\mathrm{d},j}]) \in \real^{d_{\mathrm{f}}}$, computed via~\eqref{eq:fusion} from user~$j$'s own fingerprint pair $(\tilde{\bm{r}}_j, \tilde{\bm{d}}_j)$.
A linear projection maps each user's fused representation to the multi-user feature space, $\bm{u}_j = \bm{W}_{\mathrm{u}} \bm{h}_j + \bm{b}_{\mathrm{u}} \in \real^{d_{\mathrm{u}}}$.
A scaled dot-product self-attention layer~\cite{vaswani2017attention} exchanges information
\begin{equation}
\label{eq:multi_user}
\bm{u}'_j = \bm{u}_j + \bm{W}_{\mathrm{V}}\bm{U}^{\top}\softmax\!\left(\frac{\bm{U}\bm{W}_{\mathrm{K}}^{\top}\bm{W}_{\mathrm{Q}}\bm{u}_j}{\sqrt{d_{\mathrm{u}}}}\right),
\end{equation}
where $\bm{U} = [\bm{u}_1, \ldots, \bm{u}_{K_{\mathrm{u}}}]^{\top} \in \real^{K_{\mathrm{u}} \times d_{\mathrm{u}}}$ stacks all user features, $d_{\mathrm{u}}$ is the multi-user feature dimension, and $\bm{W}_{\mathrm{Q}}, \bm{W}_{\mathrm{K}}, \bm{W}_{\mathrm{V}} \in \real^{d_{\mathrm{u}} \times d_{\mathrm{u}}}$ are the learned query, key, and value projections.
The residual connection preserves each user's individual information, while the attention mechanism allows users to share spatial context.
Layer normalization is applied after the residual, followed by a shared per-user position head $\hat{\bm{q}}_j = \bm{W}_{\mathrm{out}} \bm{u}'_j + \bm{b}_{\mathrm{out}}$, where $\bm{W}_{\mathrm{out}} \in \real^{2 \times d_{\mathrm{u}}}$ and $\bm{b}_{\mathrm{out}} \in \real^2$.
Because each $\bm{u}_j$ encodes the common propagation structure through the \ac{DT}, the attention mechanism operates on \ac{DT}-mediated representations rather than spatial proximity, implying that the multi-user improvement should be relatively insensitive to physical inter-user distance, as validated in Section~\ref{sec:exp_multiuser}.
Section~\ref{sec:exp_multiuser} emulates user groups from single-user measurements, since every \ac{UE} receives the same downlink sweep with its own receive beams and co-present \acp{UE} change each other's mean fingerprint only through mutual blockage.

\subsection{Inference Procedure}
\label{sec:inference}

At deployment, the \ac{UE} position is unknown, so the corresponding \ac{DT} fingerprint $\tilde{\bm{d}}$ cannot be queried directly at the true position.
We adopt a two-stage inference procedure that queries the \ac{DT} model after obtaining a coarse position estimate. In our implementation, the ray-traced \ac{DT} responses are cached on a dense spatial grid $\mathcal{G} \subset \mathcal{A}$ as $\{\bm{d}(\bm{q}_i)\}_{\bm{q}_i \in \mathcal{G}}$ for computational efficiency.

\textbf{Stage~1 (Coarse localization).}
Given the observed real fingerprint $\tilde{\bm{r}}$, a lightweight \ac{kNN} search over the \emph{real} fingerprint database identifies a coarse position estimate~$\hat{\bm{q}}_0$.
This step uses the same \acs{kNN} Real baseline from Section~\ref{sec:exp_baselines} and requires no neural network evaluation.
\ac{kNN} needs no training and uses only the real fingerprint. Any coarse estimator that meets the assumptions of Proposition~\ref{prop:pipeline_crb} with error covariance $\bm{C}_0' \preceq \bm{C}_0$ can replace it, and by part~(b) its pipeline bound is no larger.

\textbf{Stage~2 (\acs{DTCond} refinement).}
The \ac{DT} fingerprint at the coarse estimate, $\bm{d}(\hat{\bm{q}}_0)$, is obtained by querying the \ac{DT} model, or equivalently by reading the cached ray-traced value at the nearest grid point in our implementation, then preprocessed following Section~\ref{sec:system_preprocessing} and paired with $\tilde{\bm{r}}$.
\acs{DTCond} then produces the refined estimate $(\hat{\bm{q}},\, \bm{\sigma}) = f_{\bm{\theta}}(\tilde{\bm{r}},\, \tilde{\bm{d}}(\hat{\bm{q}}_0))$.

Because the \ac{DT} fingerprint varies smoothly in space, an error $\Delta\bm{q} = \hat{\bm{q}}_0 - \bm{q}$ gives $\bm{d}(\hat{\bm{q}}_0) \approx \bm{d}(\bm{q}) + \bm{G}\Delta\bm{q}$ (Proposition~\ref{prop:pipeline_crb}).
To make \acs{DTCond} robust to this perturbation, we augment training by randomly substituting a nearby \ac{DT} grid neighbor as the \ac{DT} input; the augmentation probability and neighborhood size are specified in Section~\ref{sec:exp_ablation}.

\section{Twin-in-the-Loop Progressive Calibration and Localization-Driven Network Configuration}
\label{sec:calibration}

\subsection{Calibration Framework}
\label{sec:calibration_framework}

The twin-in-the-loop framework, outlined in Algorithm~\ref{alg:calibration}, progressively improves localization by iterating between data collection and model training.
The \ac{DT} provides dense spatial coverage while real measurements provide calibration anchors, surveyed positions at which the fingerprint is measured and the ground-truth position is known.
As the calibration set $\mathcal{C}$ grows, the per-domain normalization statistics~\eqref{eq:normalize} stabilize, and the localization error decreases (Section~\ref{sec:exp_calibration}).

\begin{algorithm}[t]
\caption{Twin-in-the-Loop Progressive Calibration and Localization-Driven Network Configuration}
\label{alg:calibration}
\begin{algorithmic}[1]
\REQUIRE Ray-traced $\bm{d}_0$ on $\mathcal{G}$, anchor counts $\{N_{\mathrm{c},t}\}_{t=0}^{T_{\mathrm{c}}}$, $N_{\mathrm{s}}$
\STATE Survey $N_{\mathrm{c},0}$ initial anchors by~\eqref{eq:maximin} and fit $\bm{\Sigma}_{\delta,m}$
\FOR{periods $t = 0, 1, \ldots$}
\STATE \emph{Update} $\hat{\bm{\delta}}_t$ by~\eqref{eq:loop_update} and set $\bm{d}_t = \bm{d}_0 + \hat{\bm{\delta}}_t$
\STATE If anchors were added, retrain \acs{DTCond} by~\eqref{eq:normalize}--\eqref{eq:nll_loss}
\FOR{each user with a full sweep in period $t$}
\STATE \emph{Localize} $(\hat{\bm{q}}, \bm{\sigma})$; keep the sweep as a report at $\hat{\bm{q}}$
\STATE \emph{Configure} $\mathcal{B}_m$ and $\hat{m}$ by~\eqref{eq:loop_config} until the next sweep
\ENDFOR
\STATE If $t < T_{\mathrm{c}}$, add $N_{\mathrm{c},t+1} - N_{\mathrm{c},t}$ anchors by~\eqref{eq:maximin}
\ENDFOR
\end{algorithmic}
\end{algorithm}

\subsection{Progressive Data Collection}
\label{sec:calibration_progressive}

The progressive calibration analysis varies $|\mathcal{C}|$ from 5\% to 100\% of the available training data.
At each fraction, we subsample the real training data, recompute the per-domain normalization statistics~\eqref{eq:normalize} from only the available $N_{\mathrm{c}}$ positions, retrain DTCond from scratch, and evaluate on a fixed test set.
The \ac{DT} model remains available throughout, so only \emph{real} measurements are limited.

\subsection{Spatially Diverse Measurement Selection}
\label{sec:calibration_spatial}

Given a budget of $N_{\mathrm{c}}$ calibration measurements, the selection of \emph{where} to measure significantly impacts localization quality.
We adopt a maximin criterion that maximizes coverage
\begin{equation}
\label{eq:maximin}
\bm{q}_{\mathrm{new}} = \argmax_{\bm{q} \in \mathcal{A} \setminus \{\bm{q}_k\}_{k=1}^{N_{\mathrm{c}}}} \; \min_{k \in \{1, \ldots, N_{\mathrm{c}}\}} \|\bm{q} - \bm{q}_k\|.
\end{equation}
Section~\ref{sec:exp_calibration} compares this \emph{spatially diverse} selection with \emph{random} selection and \emph{uncertainty-guided} selection of the positions with the highest predicted uncertainty $\|\bm{\sigma}\|$ of a \acs{DTCond} model trained on a seed set.

\subsection{Localization-Driven Network Configuration}
\label{sec:loop_update}
\label{sec:loop_config}

\looseness=-1 Algorithm~\ref{alg:calibration} closes two loops through localization. Calibration anchors retrain \acs{DTCond} and improve the localizer (Section~\ref{sec:exp_calibration}), and its position estimates drive the network configuration and place the reports that refine the \ac{DT}, a use case that we demonstrate for beam and serving-\ac{AP} selection, two location-aware services of Section~\ref{sec:intro}. A full sweep of all 81 pairs of every \ac{AP} reveals the strongest pairs only at its own instant. Between sweeps, the network must predict them for the moving \ac{UE}. The prediction uses the tracked estimate, i.e., the position estimate of the last sweep advanced by odometry, and the \ac{RSS} there, given by the \ac{DT}. Selection compares absolute powers, but the ray-traced \ac{DT} $\bm{d}_0 \equiv \bm{d}$ ranks the strongest measured pair first at only 27.5\% of positions. Selection thus uses the updated \ac{DT} $\bm{d}_t = \bm{d}_0 + \hat{\bm{\delta}}_t$ of the twin update~\eqref{eq:loop_update}, whereas \acs{DTCond} keeps $\bm{d}_0$, since localization needs only its spatial structure (Remark~\ref{rem:spatial}). In the loop of Algorithm~\ref{alg:calibration}, the \ac{UE} is \emph{localized}, $\bm{d}_t$ at the tracked estimate \emph{configures} the pairs to train, and the next sweep, placed at the position estimate, \emph{updates} $\bm{d}_t$, so localization closes the loop.

\looseness=-1 The gap over the grid, $\bm{\delta}_{m,n} \in \real^{|\mathcal{G}|}$ for \ac{AP}~$m$ and beam pair~$n$, has the prior $\mathcal{N}(\bm{0}, \bm{\Sigma}_{\delta,m})$, $\bm{\Sigma}_{\delta,m} = \sigma_{\bar{\delta}}^2\bm{1}\bm{1}^{\top} + s_m^2\bm{K}_m$, whose two terms model the spatial mean and the position-dependent part of the gap (Remark~\ref{rem:spatial}), with a correlation matrix $\bm{K}_m \succ \bm{0}$ that decays with distance.
Observation~$k$ is a calibration anchor or a measurement report, i.e., the full sweep of a served user localized at $\hat{\bm{q}}_k$ with covariance $\hat{\bm{\Sigma}}_k = \diag(\sigma_{x,k}^2, \sigma_{y,k}^2)$. It is assigned to the node $\bm{q}_{i_k}$ nearest to its surveyed or estimated position and gives $z_{k,n} = r_{k,n} - d_{0,m,n}(\bm{q}_{i_k}) = \delta_{m,n}(\bm{q}_{i_k}) + \omega_{k,n}$, where $d_{t,m,n}$ is the $n$th entry of $\bm{d}_t^{(m)}$ and $\omega_{k,n}$ is noise of variance $\varsigma_{k,n}^2$. For an anchor, $\varsigma_{k,n}^2 = \varsigma_m^2$, the single-snapshot noise variance of \ac{AP}~$m$. For a report, $\varsigma_{k,n}^2 = \varsigma_m^2 + \bm{g}_{k,n}^{\top}(\hat{\bm{\Sigma}}_k + \bm{o}_k\bm{o}_k^{\top})\bm{g}_{k,n}$, with $\bm{g}_{k,n} = \partial d_{t,m,n}(\bm{q}_{i_k})/\partial\bm{q}$ and $\bm{o}_k = \hat{\bm{q}}_k - \bm{q}_{i_k}$. Localization uncertainty thus down-weights each report, as $\bm{C}_0$ does in~\eqref{eq:sigma_eff}. The twin update is the \ac{LMMSE} estimate, i.e., Gaussian-process regression (kriging),
\begin{equation}
\label{eq:loop_update}
\hat{\bm{\delta}}_{t,m,n} = \bm{J}_{t,m,n}^{-1}\bm{\zeta}_t,\quad
\bm{J}_{t,m,n} = \bm{\Sigma}_{\delta,m}^{-1} + \diag(\bm{\pi}_t),
\end{equation}
with posterior variance $\nu_{t,m,n}(\bm{q}_i) = [\bm{J}_{t,m,n}^{-1}]_{ii}$, where $[\bm{\pi}_t]_i$ and $[\bm{\zeta}_t]_i$ sum $\varsigma_{k,n}^{-2}$ and $z_{k,n}\varsigma_{k,n}^{-2}$ over $k \in \mathcal{K}_t$ with $i_k = i$, $\mathcal{K}_t$ being the set of observations up to period~$t$.

\begin{proposition}[Monotone twin refinement]
\label{prop:loop_monotone}
If $\mathcal{K}_t \subseteq \mathcal{K}_{t+1}$ and each $\varsigma_{k,n}^2 > 0$ is fixed when observation~$k$ enters, then $\bm{J}_{t+1,m,n} \succeq \bm{J}_{t,m,n}$ and $\nu_{t+1,m,n}(\bm{q}_i) \leq \nu_{t,m,n}(\bm{q}_i)$ $\forall \bm{q}_i \in \mathcal{G}$, so the model-predicted mean-square \ac{DT} error $\nu_{t,m,n}(\bm{q}_i) + \varsigma_m^2$ of a new measurement at $\bm{q}_i$ is non-increasing in $t$.
\end{proposition}

\begin{proof}
Nesting with fixed $\varsigma_{k,n}^2$ gives $\bm{J}_{t+1,m,n} - \bm{J}_{t,m,n} = \diag(\bm{\pi}_{t+1} - \bm{\pi}_t) \succeq \bm{0}$ and $\bm{J}_{t,m,n} \succeq \bm{\Sigma}_{\delta,m}^{-1} \succ \bm{0}$, so $\bm{J}_{t+1,m,n}^{-1} \preceq \bm{J}_{t,m,n}^{-1}$, whose $(i,i)$ entries give the claim on $\nu$.
\end{proof}

\looseness=-1 Configuration reads $\bm{d}_t$ at the node $\tilde{\bm{q}} \in \mathcal{G}$ nearest to the tracked estimate,
\begin{equation}
\label{eq:loop_config}
\mathcal{B}_m = \operatorname{top}_{N_{\mathrm{s}}}\!\bigl(\bm{d}_t^{(m)}(\tilde{\bm{q}})\bigr), \qquad
\hat{m} = \argmax_{m} \max_{n \in \mathcal{B}_m} \tilde{r}_{m,n},
\end{equation}
where $\operatorname{top}_{N_{\mathrm{s}}}$ returns the indices of the $N_{\mathrm{s}}$ largest entries and $\tilde{r}_{m,n}$ is the \ac{RSS} of pair~$n$ of \ac{AP}~$m$ measured in beam training. Until the next full sweep of the user, each \ac{AP} trains only its candidate set $\mathcal{B}_m$, and \ac{AP}~$\hat{m}$ serves the user on its strongest trained pair. The action leaves the propagation of a static scene unchanged. It sets the served link, its received power, and the training overhead.

\looseness=-1 For a user at $\bm{q}$ with fingerprint $\bm{r}^{(m)}$, the beam-selection loss $\ell_m = \max_n r^{(m)}_n - \max_{n \in \mathcal{B}_m} r^{(m)}_n$ of \ac{AP}~$m$ in dB obeys
\begin{equation}
\label{eq:loop_loss}
\ell_m \leq 2\bigl\|\bm{r}^{(m)} - \bm{d}_t^{(m)}(\tilde{\bm{q}})\bigr\|_{\infty} \leq 2\bigl\|\bm{\epsilon}_t^{(m)}(\bm{q})\bigr\|_{\infty} + 2\Lambda_{m,t}\|\tilde{\bm{q}} - \bm{q}\|,
\end{equation}
because $\hat{n} = \argmax_n d_{t,m,n}(\tilde{\bm{q}}) \in \mathcal{B}_m$, so that $\ell_m \leq r^{(m)}_{n^\star} - r^{(m)}_{\hat{n}}$ with $n^\star = \argmax_n r^{(m)}_n$, and $d_{t,m,n^\star}(\tilde{\bm{q}}) \leq d_{t,m,\hat{n}}(\tilde{\bm{q}})$. Here $\bm{\epsilon}_t^{(m)}(\bm{q}) = \bm{r}^{(m)} - \bm{d}_t^{(m)}(\bm{q})$ is the \ac{DT} error and $\Lambda_{m,t}$ satisfies $\|\bm{d}_t^{(m)}(\tilde{\bm{q}}) - \bm{d}_t^{(m)}(\bm{q})\|_{\infty} \leq \Lambda_{m,t}\|\tilde{\bm{q}} - \bm{q}\|$, so the second step is the triangle inequality. By Proposition~\ref{prop:loop_monotone}, the update never increases $\sum_n (\nu_{t,m,n}(\bm{q}) + \varsigma_m^2) = \expect\|\bm{\epsilon}_t^{(m)}(\bm{q})\|_2^2 \geq \expect\|\bm{\epsilon}_t^{(m)}(\bm{q})\|_{\infty}^2$, the model-predicted bound, $\forall \bm{q} \in \mathcal{G}$. Accurate localization reduces the second term. Thus~\eqref{eq:loop_loss} ties the beam-selection loss to the \ac{DT} and localization errors.

\section{Experimental Evaluation}
\label{sec:experiments}

\subsection{Measurement Campaign and Dataset}
\label{sec:exp_setup}

\begin{figure}[!t]
\centering
\includegraphics[width=\columnwidth]{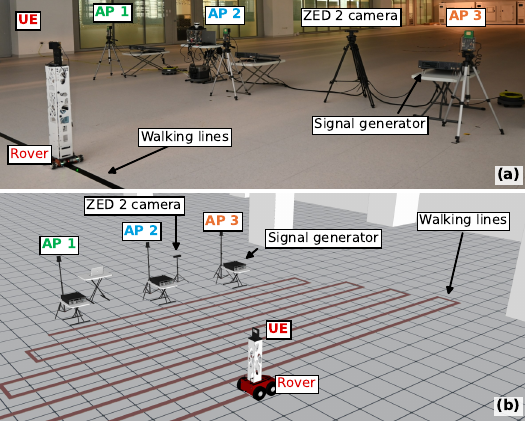}
\caption{(a)~Real indoor 28\,GHz cell-free testbed. (b)~Ray-tracing \ac{DT} reconstruction showing the 3D scene model with \ac{AP} placements and measurement trajectory.}
\label{fig:testbed}
\end{figure}

We evaluate the proposed framework on a real indoor 28\,GHz cell-free testbed, illustrated in Fig.~\ref{fig:testbed}.
The site is a static laboratory with reflecting furniture, equipment, and metal surfaces and largely unobstructed direct paths.
The testbed comprises $M = 3$ distributed \acp{AP}, each equipped with a \ac{UPA} and a codebook of $N_{\mathrm{b}} = N_{\mathrm{tx}} \times N_{\mathrm{rx}} = 9 \times 9 = 81$ beam pairs.
The coverage area is a trapezoid of approximately $9.6$\,m depth with parallel sides of $6$\,m and $19.6$\,m. 
A mobile \ac{UE} traverses $L = 13$ straight-line walking trajectories, collecting $S = 20$ repeated beam-domain measurements at each of $N_{\mathrm{p}} = 1{,}400$ positions.
The $S$ repeated measurements are averaged to form per-position fingerprints, with the per-snapshot variance used for noise estimation, yielding $\hat{\sigma}_{\mathrm{r}}^2 = 2.547$\,dB$^2$.
Ground-truth \ac{UE} positions along each trajectory were recorded using a ZED~2 stereo camera with visual-inertial odometry, providing centimeter-level positional accuracy ($\pm 2$\,cm).
The \ac{DT} is constructed using a site-specific ray-tracing propagation model in MATLAB on the same 3D scene model (Fig.~\ref{fig:testbed}), generating fingerprints at the same $N_{\mathrm{p}}$ positions.
The ray tracer uses up to one reflection, no diffraction, and a uniform wood material, and it sums the ray powers incoherently, each weighted by beam patterns fitted empirically to the arrays. The \ac{DT} is also evaluated on a grid $\mathcal{G}$ of 2{,}707 nodes spaced 0.14\,m along the walking lines and their midlines.
System parameters are summarized in Table~\ref{tab:params}.

\textbf{Domain gap statistics.}
Per-\ac{AP} Pearson correlations between real and \ac{DT} fingerprints are $\rho_1 = 0.17$, $\rho_2 = 0.34$, $\rho_3 = 0.33$, indicating a substantial domain gap. The overall real-\ac{DT} correlation computed jointly across all $M N_{\mathrm{b}} = 243$ beam elements is 0.597, higher than the per-\ac{AP} values because both domains reproduce the power differences between \acp{AP}.
The effective \ac{DT} domain gap variance is $\hat{\sigma}_{\Delta}^2 = 0.932$\,dB$^2$, smaller than the per-snapshot measurement noise $\hat{\sigma}_{\mathrm{r}}^2 = 2.547$\,dB$^2$, reflecting the \ac{DT}'s lower random variability but higher systematic bias.
The ratio $\hat{\sigma}_{\mathrm{r}}^2/\hat{\sigma}_{\Delta}^2 = 2.73$ places this testbed in the regime where \ac{DT} conditioning helps most (Remark~\ref{rem:crb_quantify}).

\begin{table}[t!]
\centering
\caption{System and Hyperparameter Configuration}
\label{tab:params}
\setlength{\tabcolsep}{2pt}
\footnotesize
\renewcommand{\arraystretch}{0.95}
\begin{tabular}{@{}lc|lc@{}}
\toprule
\textbf{Parameter} & \textbf{Value} & \textbf{Parameter} & \textbf{Value} \\
\midrule
\multicolumn{4}{@{}l}{\textit{Measurement Setup}} \\
$f_{\mathrm{c}}$ (Carrier freq.) & $28\,\mathrm{GHz}$ & BW (Bandwidth) & $40\,\mathrm{MHz}$ \\
$P_{\mathrm{tx}}$ (Transmit power) & $10\,\mathrm{dBm}$ & Noise floor & $-110\,\mathrm{dBm}$ \\
$M$ (No.\ of APs) & 3 & UE height & $1.0\,\mathrm{m}$ \\
$N_{\mathrm{b}}$ (Beam pairs/AP) & $9{\times}9{=}81$ & $MN_{\mathrm{b}}$ (FP dim.) & 243 \\
$N_{\mathrm{p}}$ (Meas.\ positions) & 1{,}400 & $L$ (Trajectories) & 13 \\
$S$ (Snapshots/pos.) & 20 & $\hat{\sigma}_{\mathrm{r}}^2$ (Noise var.) & $2.547\,\mathrm{dB}^2$ \\
$\hat{\sigma}_{\Delta}^2$ (DT gap var.) & $0.932\,\mathrm{dB}^2$ & $\rho_m$ (AP corr.) & 0.17, 0.34, 0.33 \\
Train/val/test split & 70/10/20 & Split / eval.\ seeds & 42 / 5 \\
\midrule
\multicolumn{4}{@{}l}{\textit{DTCond Hyperparameters}} \\
$d_{\mathrm{e}}$ (Embed.\ dim.) & 32 & $d_{\mathrm{h}}$ (Hidden dim.) & 64 \\
$d_{\mathrm{f}}$ (Fusion dim.) & 64 & $d_{\mathrm{a}}$ (Attn.\ dim.) & 32 \\
$d_{\mathrm{u}}$ (Multi-user dim.) & 64 & $p_{\mathrm{d}}$ (Dropout) & 0.1 \\
$\nabla_{\max}$ (Grad.\ clip) & 5.0 & Batch size & 32 \\
\bottomrule
\end{tabular}
\\[4pt]
\footnotesize Training uses Adam (initial learning rate $10^{-3}$, weight decay $10^{-5}$) with cosine annealing to $10^{-6}$, 60-epoch MSE warmup then 90-epoch NLL, early stopping (patience 25).
\end{table}
\vspace{-0.35mm}
\subsection{Baselines and Metrics}
\label{sec:exp_baselines}

We compare \acs{DTCond} with nine baselines of three types.

\begin{itemize}
\item \textbf{Non-parametric baselines.}
(1)~\textbf{\acs{kNN} Real.} \ac{kNN} with $k\!=\!5$ nearest neighbors using real fingerprints only ($k \in \{3,\ldots,7\}$ is the conventional range for fingerprint localization~\cite{torres2015fingerprint}).
(2)~\textbf{\acs{kNN} DT.} \ac{kNN} using \ac{DT} fingerprints only.
(3)~\textbf{\acs{kNN} Real+DT.} \ac{kNN} with concatenated real and \ac{DT} fingerprints.

\item \textbf{Neural network baselines.}
(4)~\textbf{NN RealOnly.} \ac{MLP} trained on real fingerprints only.
(5)~\textbf{NN Transfer.} Pre-trained on \ac{DT} data, fine-tuned on real data with all layers unfrozen at a reduced learning rate ($10^{-4}$). Transfer learning underperforms because the two domains differ in their underlying observation processes, not merely their distributions; ray-traced features encode propagation artifacts that limited fine-tuning data cannot correct, consistent with the \ac{DA} failures below.
(6)~\textbf{NN Concat.} \ac{MLP} with concatenated real and \ac{DT} input (same total parameters as \acs{DTCond}).

\item \textbf{\ac{DA} baselines.}
(7)~\textbf{NN \acs{DANN}.} \ac{DANN}~\cite{ganin2016} with gradient reversal for domain-invariant features.
(8)~\textbf{NN \acs{CORAL}.} Deep \ac{CORAL}~\cite{sun2016coral} minimizing domain covariance discrepancy.
(9)~\textbf{NN \acs{MMD}.} Shared encoder with Gaussian RBF kernel \ac{MMD} penalty~\cite{gretton2012mmd} (median heuristic bandwidth). For all three \ac{DA} methods, the penalty weight (\ac{DANN} $\lambda$, \ac{CORAL} weight, \ac{MMD} weight) was swept over $\{0.1, 1, 10, 100\}$; best values are reported.
\end{itemize}

The primary metric is mean Euclidean error (m) on the 280-position test set.
We also report median error, 90th percentile error (P90), and standard deviation across five random seeds for neural methods.

\subsection{Localization Accuracy Comparison}
\label{sec:exp_ablation}

\begin{figure}[!t]
\centering
\includegraphics[width=0.9\columnwidth]{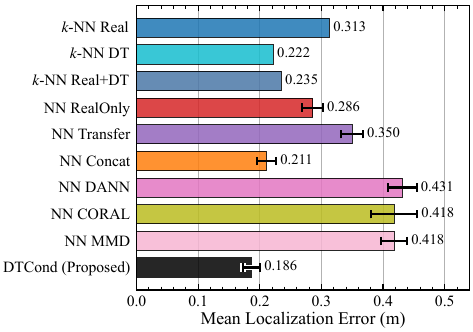}
\caption{Ten-method ablation (five seeds, mean $\pm$ std).}
\label{fig:ablation}
\end{figure}

Fig.~\ref{fig:ablation} presents the ablation. Three key findings emerge.
\begin{itemize}
\item \textbf{DT conditioning is the main source of gain.}
\looseness=-1
All \ac{DT}-conditioned methods outperform the real-only baselines (25--41\% over \acs{kNN} Real). \acs{DTCond}, NN~Concat, \acs{kNN} Real+DT, and \acs{kNN} DT achieve $0.186 \pm 0.014$\,m, $0.211 \pm 0.016$\,m, 0.235\,m, and 0.222\,m, respectively, versus $0.286 \pm 0.017$\,m for NN~RealOnly and 0.313\,m for \acs{kNN} Real ($\pm$ values denote std across five seeds; \ac{kNN} methods are deterministic). Notably, simple concatenation (\acs{kNN} Real+DT) already yields 25\% improvement, and \acs{DTCond} reaches 41\%. All gains occur despite low per-\ac{AP} correlations ($\rho_m = 0.17$--$0.34$), confirming that the \ac{DT} contributes through spatial structure rather than absolute fidelity.
The 12\% lead of \acs{DTCond} over NN~Concat on this split (paired $t$-test across five seeds, $p$-value $= 0.022$) is consistent with the heteroscedastic refinement of Proposition~\ref{prop:het_crb}: NN~Concat has no explicit per-\ac{AP} weighting, whereas the \ac{AP} attention of \acs{DTCond} can weight each \ac{AP} by its local precision, as the heteroscedastic \ac{FIM} does. The lead persists under \ac{LOLO} ($0.257$ vs.\ $0.264$\,m, $p$-value $= 0.042$), and the precision-aware variant of Section~\ref{sec:exp_lolo} extends it to $0.244$\,m.
All methods share the split and preprocessing, and \acs{kNN} Real+DT and NN~Concat differ from their real-only counterparts only in the \ac{DT} input, so their 25\% and 26\% gains come from the paired input and the architecture adds the 12\% above.
The \ac{DT}-conditioned results of the ablation and of all later sections read the \ac{DT} fingerprint at the true position, as the oracle \ac{CRB} assumes, unless marked as deployable.

\item \textbf{\ac{DA} baselines underperform.}
\looseness=-1
\ac{DANN}, \ac{CORAL}, and \ac{MMD} achieve $0.431 \pm 0.023$\,m, $0.418 \pm 0.038$\,m, and $0.418 \pm 0.021$\,m, roughly $2\times$ worse than \acs{DTCond} and worse than \acs{kNN} Real ($p$-value $< 0.01$, paired $t$-test).
NN~Transfer also degrades performance ($0.350 \pm 0.018$\,m vs.\ $0.286 \pm 0.017$\,m for NN~RealOnly).
Forcing the two domains into a shared representation suppresses complementary information; the fundamental mismatch in observation processes---not a simple distribution shift---explains why alignment-based approaches suppress information that conditioning preserves.

\item \textbf{Deployable pipeline performance.}
The two-stage pipeline (Section~\ref{sec:inference}) with augmented training (probability 0.3, eight \ac{DT} grid neighbors) achieves 0.284\,m against 0.313\,m for \ac{kNN} Real and 0.345\,m for the same pipeline without augmentation.
The gap from the non-augmented oracle (0.186\,m) decomposes into two quantifiable components: (i)~an augmentation cost ($0.186 \to 0.232$\,m), necessary to tolerate retrieval noise, and (ii)~retrieval mismatch ($0.232 \to 0.284$\,m), predicted by Proposition~\ref{prop:pipeline_crb}, whose diagnostic ratio $\|\bm{G}\bm{J}_{\mathrm{r}}^{-1}(\bm{q})\bm{G}^{\top}\|/\sigma_{\Delta}^2$ averages~3.21.
Improving the coarse stage would yield the largest marginal return (Proposition~\ref{prop:pipeline_crb}).
\end{itemize}

\subsection{\acs{CRB} Validation}
\label{sec:exp_crb}

\begin{table}[!htbp]
\centering
\caption{Oracle \ac{CRB} Analysis on Test Positions ($n = 280$)}
\label{tab:crb}
\footnotesize
\renewcommand{\arraystretch}{0.95}
\begin{tabular}{@{}lccc@{}}
\toprule
& \textbf{Real-Only} & \textbf{Joint (Real+DT)} & \textbf{Reduction} \\
\midrule
Mean \acs{CRB} (m) & $0.274 \pm 0.072$ & $0.178 \pm 0.060$ & 34.9\% \\
Median \acs{CRB} (m) & 0.277 & 0.172 & 37.9\% \\
P10 \acs{CRB} (m) & 0.180 & 0.107 & 40.6\% \\
P90 \acs{CRB} (m) & 0.364 & 0.261 & 28.3\% \\
\bottomrule
\end{tabular}
\end{table}

Table~\ref{tab:crb} validates the oracle \ac{CRB} analysis from Section~\ref{sec:system_crb}.
Fingerprint Jacobians are computed numerically at each of the 280 test positions via least-squares estimation from ten nearest positions, with $\hat{\sigma}_{\mathrm{r}}^2 = 2.547$\,dB$^2$ and $\hat{\sigma}_{\Delta}^2 = 0.932$\,dB$^2$.

The oracle joint \ac{CRB} (mean 0.178\,m) is 34.9\% lower than the real-only \ac{CRB} (0.274\,m), confirming that the \ac{DT} contains additional spatial information despite the low per-AP correlations in Table~\ref{tab:params}.
The improvement is robust to the Jacobian estimation neighborhood size (6, 8, 10, or 12 nearest positions; mean reduction varies by less than $\pm 1.5\%$).

\textbf{Robustness to model misspecification.}
A Kolmogorov--Smirnov test rejects exact Gaussianity ($p$-value ${< 10^{-9}}$), though departures are mild (skewness $< 0.21$, excess kurtosis $\approx -0.2$).
The misspecified \ac{CRB}~\cite{richmond2015misspecified} changes the joint improvement of the Gaussian \ac{CRB} by only 0.3 percentage points, confirming robustness to distributional assumptions.

\begin{figure}[!t]
\centering
\includegraphics[width=0.8\columnwidth]{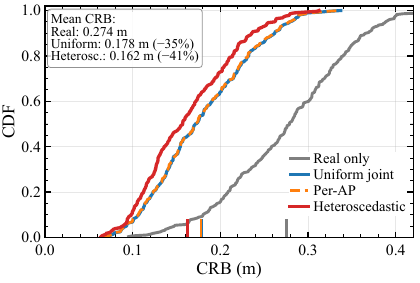}
\caption{\Ac{CDF} of per-position oracle CRB on 280 test positions under four variance models. The local heteroscedastic bound is tighter than the uniform bound at 97.1\% of positions.}
\label{fig:crb_cdf}
\end{figure}

\begin{table}[!htbp]
\centering
\caption{Extended \acs{CRB} Analysis Including Pipeline Bound ($n = 280$)}
\label{tab:crb_extended}
\footnotesize
\resizebox{\columnwidth}{!}{%
\begin{tabular}{@{}lcccc@{}}
\toprule
& \textbf{Real-Only} & \textbf{Pipeline} & \textbf{Oracle (unif.)} & \textbf{Oracle (het.)} \\
\midrule
Mean \acs{CRB} (m) & $0.274 \pm 0.072$ & $0.220 \pm 0.065$ & $0.178 \pm 0.060$ & $0.162 \pm 0.051$ \\
Median \acs{CRB} (m) & 0.277 & 0.220 & 0.172 & 0.157 \\
$\Delta$ vs.\ real & --- & 19.6\% & 34.9\% & 40.8\% \\
\midrule
Achieved error (m) & 0.313 & $0.284 \pm 0.013$ & $0.186 \pm 0.014$ & --- \\
\bottomrule
\end{tabular}%
}
\end{table}

Allowing the per-AP gap variance to vary tightens the bound to 0.162~m (40.8\% over real-only and 9.1\% over the uniform bound), and the heteroscedastic bound is tighter than the uniform bound at 97.1\% of test positions (Fig.~\ref{fig:crb_cdf}).

\textbf{Pipeline \acs{CRB} validation.}
With $\bm{C}_0 = \bm{J}_{\mathrm{r}}^{-1}(\bm{q})$, the mean pipeline \ac{CRB} is 0.220\,m (Table~\ref{tab:crb_extended}), satisfying the sandwich inequality~\eqref{eq:sandwich} at all 280 test positions. The achieved pipeline error (0.284\,m) exceeds the bound by 0.064\,m, reflecting the \ac{kNN} coarse stage, finite data, and architecture capacity; the ratio $\|\bm{G}\bm{J}_{\mathrm{r}}^{-1}(\bm{q})\bm{G}^{\top}\|/\sigma_{\Delta}^2$ averages 3.21, confirming retrieval mismatch as a substantial driver of the oracle-to-pipeline gap.
With $\bm{C}_0$ set to the empirical error covariance of the \ac{kNN} coarse stage, the mean pipeline bound is 0.231\,m. A more accurate coarse stage lowers the bound toward the oracle 0.178\,m (Proposition~\ref{prop:pipeline_crb}(b)), and the achieved error falls from 0.284\,m with \ac{kNN} to 0.232\,m with an exact coarse stage ($\bm{C}_0 = \bm{0}$). An unbiased single-sweep coarse stage has $\bm{C}_0 \succeq \bm{J}_{\mathrm{r}}^{-1}$ and, by part~(b), keeps the bound $\geq 0.220$\,m.

\textbf{Per-\acs{AP} marginal \acs{DT} utility validation.}
Despite the lowest fidelity ($\rho_1 = 0.17$), \ac{AP}~1 yields the highest utility $\mu_m$ at 19\% of positions because its \ac{DT} Jacobian complements those of \acp{AP}~2--3 along the weak eigendirection of $\bm{J}_{\mathrm{r}}(\bm{q})$, which carries 73\% of the \ac{DT}'s \ac{CRB} reduction (Remark~\ref{rem:crb_directional}).

\textbf{Gradient alignment verification.}
The per-\ac{AP} Jacobian cosine similarities~\eqref{eq:jacobian_cosine},
averaged over 280 test positions, are $\bar{\gamma}_1 = 0.52$, $\bar{\gamma}_2 = 0.63$, $\bar{\gamma}_3 = 0.58$---well above the per-\ac{AP} Pearson (amplitude) correlations $\rho_m \in \{0.17, 0.34, 0.33\}$ from~\eqref{eq:correlation}, confirming spatial gradient preservation despite low absolute power agreement (Remark~\ref{rem:spatial}, Proposition~\ref{prop:crb}).

\textbf{Alignment factor distribution.}
The scalar alignment factor $\hat{c}(\bm{q}_k)$ from~\eqref{eq:c_hat} has median $0.71$ (IQR $[0.49,\, 0.92]$) across 280 test positions.
Combined with $\sigma_{\mathrm{r}}^2/\sigma_{\Delta}^2 = 2.73$, expression~\eqref{eq:crb_aligned} predicts an improvement factor of $\sqrt{1 + 0.71^2 \times 2.73} \approx 1.54$ at the median $\hat{c}$, matching the ratio $0.274/0.178 = 1.54$ of the mean bounds in Table~\ref{tab:crb}.

\subsection{Spatial Generalization}
\label{sec:exp_lolo}

\begin{figure}[!t]
\centering
\includegraphics[width=\columnwidth]{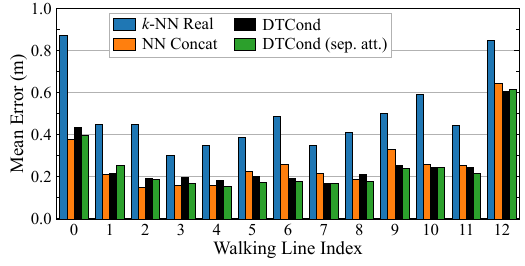}
\caption{Per-line mean localization error under \ac{LOLO} evaluation across 13 trajectories for four methods. The per-AP precision-aware variant (\acs{DTCond} sep.\ att., green) improves on baseline \acs{DTCond} (black) at 10 of 13 held-out lines.}
\label{fig:lolo}
\end{figure}

The \ac{LOLO} evaluation trains on 12 of the 13 walking lines and tests on the held-out line, rotating across all 13 folds.
\acs{DTCond} achieves $0.257 \pm 0.119$\,m mean error (95\% CI $[0.185, 0.329]$), a 48\% improvement over \ac{kNN} Real (0.495\,m; Fig.~\ref{fig:lolo}).
All \ac{DT}-conditioned methods substantially outperform real-only baselines: \acs{kNN} DT reaches 0.346\,m (30\% over \ac{kNN} Real), NN~Concat $0.264 \pm 0.126$\,m (47\%), while NN~RealOnly reaches only 0.395\,m.
\acs{DTCond} also has the lowest cross-line standard deviation (0.119 vs.\ 0.171 for \ac{kNN} Real and 0.126 for NN~Concat), the most consistent generalization.

\textbf{Heteroscedastic attention closes the LOLO gap further.} Proposition~\ref{prop:het_crb} motivates a model that expresses the per-AP precision allocation $1/\sigma_{\Delta,m}^2(\bm{q})$ of the \ac{DT} branch separately from the real branch. We evaluate a lightweight variant where the real and DT branches use separate attention heads (under 5\,k extra parameters, same training protocol). This variant attains $0.244 \pm 0.124$~m (51\% over \acs{kNN}~Real), improving baseline \acs{DTCond} on 10 of 13 held-out lines with per-line gains up to 17\% (Fig.~\ref{fig:lolo}). The random-split result is unchanged ($0.189 \pm 0.013$~m vs.\ $0.186 \pm 0.014$~m, $p$-value $= 0.73$), so precision-aware attention pays off under distribution shift.

\subsection{Robustness to DT Imperfections and Emulated Blockage}
\label{sec:exp_robust}

\begin{figure}[t!]
\centering
\begin{minipage}[t]{0.48\columnwidth}
\centering
\includegraphics[width=\linewidth]{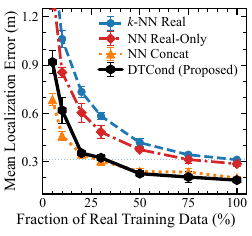}
\caption{Localization error vs.\ fraction of real training data (5--100\%).}
\label{fig:progressive}
\end{minipage}\hfill
\begin{minipage}[t]{0.48\columnwidth}
\centering
\includegraphics[width=\linewidth]{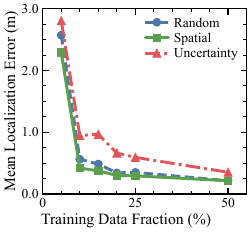}
\caption{Measurement selection strategies across budget levels.}
\label{fig:selection}
\end{minipage}
\end{figure}

\begin{table}[!t]
\centering
\caption{Mean Error (m) Under DT Imperfections (Five Seeds, DT Fingerprint at the True Position)}
\label{tab:robust}
\scriptsize
\setlength{\tabcolsep}{2pt}
\renewcommand{\arraystretch}{0.95}
\begin{tabular}{@{}lcc@{}}
\toprule
\textbf{Condition (levels)} & \textbf{\acs{DTCond}} & \textbf{\acs{kNN} Real+DT} \\
\midrule
\ac{AP} offset (0.25 / 0.5 / 1\,m) & 0.194 / 0.194 / 0.198 & 0.236 / 0.233 / 0.248 \\
\ac{AP} yaw offset ($5^{\circ}$ / $10^{\circ}$) & 0.169 / 0.191 & 0.237 / 0.239 \\
Permittivity ($-30$ / $+30\%$) & 0.213 / 0.198 & 0.236 / 0.236 \\
Conductivity $+30\%$ / concrete walls & 0.205 / 0.195 & 0.236 / 0.240 \\
Reflection order 0 / 2 / one diffraction & 0.198 / 0.195 / 0.187 & 0.233 / 0.236 / 0.235 \\
Pattern (cosine / ideal / ideal, steered) & 0.193 / 0.226 / 0.245 & 0.253 / 0.325 / 0.248 \\
Grid spacing (0.29 / 0.57 / 1.14\,m) & 0.202 / 0.203 / 0.207 & 0.254 / 0.268 / 0.264 \\
\bottomrule
\end{tabular}
\\[3pt]
\parbox{\columnwidth}{\scriptsize Both methods are retrained in every condition. Rows 1--6 re-trace the \ac{DT} with one change each (0.187 and 0.236\,m when re-traced without changes). Ideal: ideal array patterns, steered to the measured beam peaks in the last entry.}
\end{table}

\looseness=-1 Table~\ref{tab:robust} rebuilds the \ac{DT} with one modeling error at a time in geometry, materials, propagation, and antenna patterns, and thins the \ac{DT} grid. All re-traced \acp{DT} keep \acs{DTCond} within 0.169--0.245\,m, below NN~RealOnly and \ac{kNN} Real (0.286 and 0.313\,m), while the root-mean-square (RMS) real/\ac{DT} gap, net of the mean of each \ac{AP}, grows from 7.7 to 12.9\,dB. The retrained network tolerates these errors, consistent with Lemma~\ref{lem:sys_gradient}, which bounds the gap gradient, not its level.

\looseness=-1 To emulate blockage, the \ac{LoS} component of every beam pair of one random \ac{AP} per test position, obtained from the ray-traced paths and the beam patterns of the \ac{DT}, is attenuated by 20\,dB in the measured fingerprint, a typical human-body loss at 28\,GHz~\cite{3gpp38901}. The median received power falls by 6.6\,dB, and the \ac{DT} does not model the blocker. The \acs{DTCond} error grows from 0.186 to 0.280\,m, whereas \acs{kNN} Real and NN~RealOnly reach 0.730 and 1.003\,m. Blockage-aware training, with the \ac{LoS} of a random \ac{AP} attenuated by 10--30\,dB in 30\% of the samples, keeps \acs{DTCond} at 0.205\,m. Under this emulated blockage, the deployable pipeline, whose \ac{kNN} coarse stage is itself blocked, reaches 0.43\,m, well below both real-only methods.

\subsection{Progressive Calibration and Data Efficiency}
\label{sec:exp_calibration}

Fig.~\ref{fig:progressive} plots localization error against the fraction of real training data (five seeds per method per fraction).

\textbf{50\% measurement reduction.}
\acs{DTCond} at 50\% (0.226 $\pm$ 0.019\,m) outperforms \ac{kNN} Real at 100\% (0.313\,m) by 28\%, effectively halving the measurement budget.
NN~Concat tracks \acs{DTCond} closely above 25\% data (Fig.~\ref{fig:progressive}), confirming that \ac{DT} conditioning---not attention alone---drives the gain.

\textbf{Low-data crossover.}
Below 20\% data, NN~Concat's simpler architecture converges faster and outperforms \acs{DTCond} (Fig.~\ref{fig:progressive}); the learnable \ac{AP} attention mechanism requires sufficient spatial coverage ($\geq$20\%) to learn meaningful per-\ac{AP} weights.
At 5\% data (49 positions), \acs{DTCond} achieves 0.919\,m versus 1.614\,m for \ac{kNN} Real (43\% improvement), but NN~Concat reaches 0.69\,m at the same fraction.
The \ac{DT}'s advantage over real-only methods peaks at 20--50\% data.

Fig.~\ref{fig:selection} compares measurement selection strategies using \ac{kNN} Real+DT to isolate the selection criterion from the neural network's learning dynamics.
At 25\% budget, spatially diverse selection achieves 0.293\,m versus 0.350\,m for random (16\% better) and 0.593\,m for uncertainty-guided; the advantage grows at smaller budgets, confirming that \emph{spatial coverage should be the primary criterion} for calibration deployment.

\subsection{Closed-Loop Localization-Driven Configuration}
\label{sec:exp_loop}

\begin{figure}[!t]
\centering
\includegraphics[width=\columnwidth]{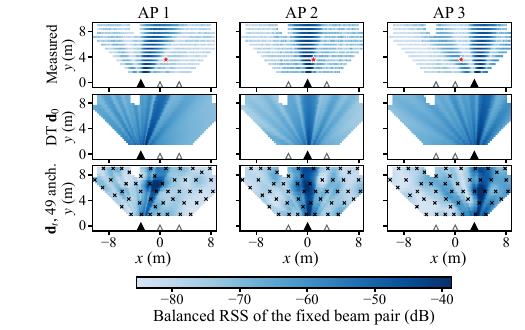}
\caption{Power-balanced \ac{RSS} (Section~\ref{sec:system_preprocessing}) of the fixed beam pair of highest mean measured power at each \ac{AP} (columns), measured (top) and predicted by $\bm{d}_0$ (middle) and by $\bm{d}_t$ updated from 49 calibration anchors (crosses, bottom), an update that reduces the RMS difference at the test positions by 19--54\% per \ac{AP}. Triangles mark the \acp{AP} (filled: \ac{AP} of the column) and the star the position of Fig.~\ref{fig:beam_heatmaps}.}
\label{fig:rss_maps}
\end{figure}

\begin{figure}[!t]
\centering
\includegraphics[width=\columnwidth]{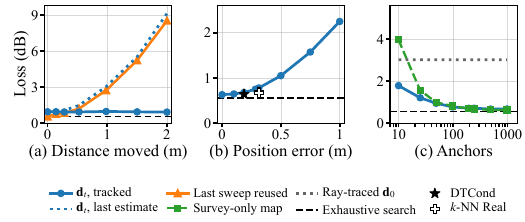}
\caption{Beam-selection loss, averaged over \acp{AP}, versus (a)~the distance moved since the last full sweep (49 anchors), (b)~the mean position error (977 anchors), and (c)~the number of anchors. In (a), $\bm{d}_t$ is read at the \ac{kNN} Real estimate of the sweep, advanced by the recorded displacement (tracked) or not (last estimate). In (b), Gaussian errors perturb the true positions, and markers give the estimates of \acs{DTCond} and \ac{kNN} Real. In (c), the survey-only map is kriged from the anchors without the \ac{DT}.}
\label{fig:closed_loop}
\end{figure}

\looseness=-1 \textbf{Closed-loop replay.} We replay the loop on the 280 test positions with $N_{\mathrm{s}} = 3$ of the 81 pairs of each \ac{AP}. The replay is valid because the action leaves the propagation unchanged and each recorded sweep holds all 81 pairs of every \ac{AP}. The metric extends $\ell_m$ of~\eqref{eq:loop_loss} to beam training and averages it over \acp{AP} and positions. Each \ac{AP} selects the pair of $\mathcal{B}_m$ on one snapshot and is scored on the 20-snapshot mean, so even exhaustive search loses 0.55\,dB to snapshot noise. Unless noted, \ac{kNN} Real, the first stage of the deployable pipeline, localizes the \ac{UE} from the 977 training positions. The anchors of $\bm{d}_t$ are nested maximin subsets~\eqref{eq:maximin} of them, and $\sigma_{\bar{\delta}}^2 = 10^4$\,dB$^2$. In Fig.~\ref{fig:rss_maps}, $\bm{d}_0$ places the main-beam region of each \ac{AP} where the measurements do (Remark~\ref{rem:spatial}) but with a level offset that the update reduces.

\looseness=-1 \textbf{Prediction between sweeps.} At the sweep position, the three strongest swept pairs match exhaustive search. Reusing them as the \ac{UE} moves loses 0.85, 2.74, and 8.54\,dB after 0.28, 0.98, and 2.0\,m (Fig.~\ref{fig:closed_loop}(a)). At the tracked estimate, $\bm{d}_t$ stays at 0.64--0.72\,dB with 977 anchors and 0.93--0.98\,dB with 49. At the unadvanced last estimate, it degrades as the last sweep does, so the gain comes from the tracked position. With 49 anchors, the link served by the \ac{AP} of the strongest trained pair loses 0.81--0.93\,dB, against 2.73\,dB after 1\,m for the last sweep. For a 1\,dB target, the last sweep must be repeated every 0.38\,m, whereas $\bm{d}_t$ with 49 anchors needs $\geq 5\times$ fewer full sweeps.

\looseness=-1 \textbf{Localization accuracy.} Fig.~\ref{fig:closed_loop}(b) isolates the localization term of~\eqref{eq:loop_loss}. With 977 anchors, the loss grows from 0.63\,dB at true positions to 0.78, 1.05, and 2.25\,dB for mean errors of 0.31, 0.5, and 1\,m. The three pairs per \ac{AP} most often strongest at the anchors lose 6.5\,dB. Mean errors below 0.47\,m, such as the 0.284\,m of the deployable pipeline, keep the loss below 1\,dB. The \acs{DTCond} and \ac{kNN} Real estimates lose 0.64 and 0.68\,dB.

\looseness=-1 \textbf{Value of the \ac{DT}.} At the sweep position, $\bm{d}_t$ loses 1.8, 0.95, and 0.68\,dB with 10, 49, and 977 anchors, against 3.0\,dB for $\bm{d}_0$ (Fig.~\ref{fig:closed_loop}(c)). With 10 anchors, the survey-only map loses 4.0\,dB, more than $\bm{d}_0$, because only $\bm{d}_t$ carries the ray-traced structure that bridges sparse anchors.

\looseness=-1 \textbf{Feedback through localization.} \ac{kNN} Real and $\bm{d}_t$ use 98 anchors. The sweeps of served \acp{UE} at 1,159 further positions, each scored before it enters $\bm{d}_t$, are placed at their \ac{kNN} Real estimates (0.94\,m mean error, validation RMS error as $\bm{\sigma}$) without ground truth. They lower the loss from 1.48 to 1.32\,dB, on par with 1.39\,dB at true positions.

\subsection{Multi-User Localization (Emulated Users)}
\label{sec:exp_multiuser}

We evaluate multi-user localization (Section~\ref{sec:framework_multiuser}) by randomly grouping positions to emulate $K_{\mathrm{u}}$ simultaneous users, training with 2{,}000 random groups per epoch.
The independent baseline ($K_{\mathrm{u}}\!=\!1$, 0.232\,m) differs from the main ablation (0.186\,m) due to this protocol.
Positions are partitioned into training, validation, and test sets before grouping, and each group lies in one partition.
Cross-user self-attention~\eqref{eq:multi_user} yields $K_{\mathrm{u}}\!=\!2$: 0.160\,m ($\pm$0.022, 31\%), $K_{\mathrm{u}}\!=\!3$: 0.141\,m ($\pm$0.013, 39.5\%), and $K_{\mathrm{u}}\!=\!4$: 0.138\,m ($\pm$0.012, 40.5\%), confirming saturation beyond three simultaneous users.

\begin{figure}[!t]
\centering
\begin{minipage}[t]{0.48\columnwidth}
\centering
\includegraphics[width=\linewidth]{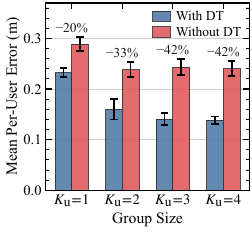}
\caption{Multi-user \ac{DT} ablation across group sizes $K_{\mathrm{u}} = 1$--$4$.}
\label{fig:multiuser_ablation}
\end{minipage}\hfill
\begin{minipage}[t]{0.48\columnwidth}
\centering
\includegraphics[width=\linewidth]{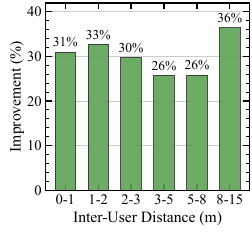}
\caption{Per-user improvement vs.\ inter-user distance ($K_{\mathrm{u}} = 2$, 1{,}971 pairs).}
\label{fig:multiuser_distance}
\end{minipage}
\end{figure}

\textbf{DT-mediated mechanism ablation.}
Without \ac{DT} conditioning, multi-user grouping yields only 17.5\% at $K_\mathrm{u} = 2$ (vs.\ 31.0\%) and 15.7\% at $K_\mathrm{u} = 3$ (vs.\ 39.5\%), confirming that the \ac{DT} raises the multi-user benefit $1.8$-fold at $K_\mathrm{u}\!=\!2$ and $2.5$-fold at $K_\mathrm{u}\!=\!3$ (Fig.~\ref{fig:multiuser_ablation}).

\textbf{Distance independence.}
Stratifying by inter-user distance across 1{,}971 pairs shows gains of 26--36\% in every bin (Fig.~\ref{fig:multiuser_distance}), so the coupling does not decay with distance, as proximity-based coupling would.
Attention weights correlate strongly with \ac{DT} fingerprint cosine similarity ($r = 0.640$, $r_s = 0.763$, both $p$-values $< 10^{-4}$), so users attend to one another through their \ac{DT} fingerprints.

\subsection{Trajectory Quality}
\label{sec:exp_trajectory}

We evaluate trajectory reconstruction fidelity on the $L = 13$ walking trajectories via the \emph{path length error} (\%)
\begin{equation}
\label{eq:path_error}
\epsilon_{\mathrm{path}} = \frac{|\ell_{\mathrm{est}} - \ell_{\mathrm{true}}|}{\ell_{\mathrm{true}}} \times 100\%,
\end{equation}
where $T$ is the number of positions along the trajectory, $\ell_{\mathrm{est}} = \sum_{t=1}^{T-1} \|\hat{\bm{q}}_{t+1} - \hat{\bm{q}}_t\|$, and $\ell_{\mathrm{true}} = \sum_{t=1}^{T-1} \|\bm{q}_{t+1} - \bm{q}_t\|$ are the estimated and true cumulative path lengths.
We also report the \emph{smoothness} metric $\xi$, defined as the mean acceleration magnitude of the estimated trajectory
\begin{equation}
\label{eq:smoothness}
\xi = \frac{1}{T\!-\!2}\sum_{t=1}^{T-2} \|\hat{\bm{q}}_{t+2} - 2\hat{\bm{q}}_{t+1} + \hat{\bm{q}}_t\|.
\end{equation}

\begin{table}[!htbp]
\centering
\caption{Trajectory Reconstruction Quality (13 Trajectories)}
\label{tab:trajectory}
\footnotesize
\renewcommand{\arraystretch}{0.95}
\begin{tabular}{@{}lccc@{}}
\toprule
\textbf{Method} & \textbf{Pos.\ Err.} $\epsilon_{\mathrm{pos}}$ \textbf{(m)} & \textbf{Path Err.} $\epsilon_{\mathrm{path}}$ \textbf{(\%)} & \textbf{Smooth.} $\xi$ \\
\midrule
\acs{kNN} Real        & 0.306 & 10.2 & 0.872 \\
\acs{kNN} Real+DT     & 0.238 & 7.5  & 0.807 \\
NN RealOnly        & 0.283 & 12.4 & 0.934 \\
NN Concat           & 0.232 & 5.4  & 0.813 \\
\textbf{\acs{DTCond}} & \textbf{0.205} & \textbf{4.5}  & \textbf{0.790} \\
\bottomrule
\end{tabular}
\end{table}

Table~\ref{tab:trajectory} reports trajectory quality metrics evaluated across all $L = 13$ trajectories (1{,}400 positions) using the random-split--trained model; positional errors differ from Fig.~\ref{fig:ablation} and Table~\ref{tab:crb_extended} because evaluation spans the complete trajectories rather than the 280-position random test set, ensuring full path continuity for the path-length and smoothness metrics.
\acs{DTCond} achieves the lowest path error (4.5\%) and smoothest trajectories ($\xi = 0.790$), while also providing uncertainty estimation, \ac{AP} attention, and multi-user cooperation.

\begin{figure}[!t]
\centering
\includegraphics[width=\columnwidth]{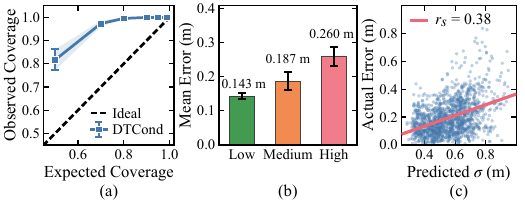}
\caption{Uncertainty reliability analysis (five seeds, 280 positions). (a)~Reliability diagram. (b)~Error by predicted uncertainty tercile. (c)~Actual error vs.\ predicted $\|\bm{\sigma}\|$.}
\label{fig:reliability}
\end{figure}

\subsection{Discussion and Design Insights}
\label{sec:exp_discussion}

\textbf{Computational cost and scalability.}
\acs{DTCond} has $\approx$32{,}000 parameters ($\mathcal{O}(d_{\mathrm{h}} N_{\mathrm{b}} + d_{\mathrm{e}} d_{\mathrm{h}} + M N_{\mathrm{b}} d_{\mathrm{a}} + d_{\mathrm{h}} d_{\mathrm{f}})$; linear in $M$) and requires 0.008\,ms per estimate on a single GPU (0.024\,ms for the two-stage pipeline), well within \ac{mmWave} beam-tracking frame rates.
On a laptop CPU (Intel Core i7-1165G7) at batch size one, including preprocessing, a query takes 0.37--0.46\,ms for the \ac{kNN} methods, 0.53--0.66\,ms for the other networks, 1.12\,ms for \acs{DTCond}, and 2.32\,ms for the pipeline, far below one beam sweep of about 180\,ms, whereas the GPU times above are amortized over batches.

\textbf{Hyperparameter robustness.}
Varying $d_{\mathrm{e}} \in \{16, 32, 64, 128\}$, dropout rate $\in \{0, 0.1, 0.2, 0.3\}$, and warmup epochs $\in \{20, 40, 60, 80\}$ one at a time (3 seeds each) keeps the mean errors within 18\% of each other.

\textbf{Uncertainty quality.}
The predicted $\|\bm{\sigma}\|$ correlates with actual error ($r_s = 0.383$, $p$-value $< 10^{-8}$) and tercile-stratified errors increase monotonically (0.143\,m, 0.187\,m, 0.260\,m), so the heteroscedastic output ranks prediction difficulty (Fig.~\ref{fig:reliability}).

\looseness=-1 \textbf{Operation below 7\,GHz.} Under the model of Section~\ref{sec:system_crb}, the bounds indicate two trends. At 3.5\,GHz, an array of fixed aperture forms beams about eight times wider, which reduces the angular information in $\bm{J}_{\mathrm{r}}$ and $\bm{J}_{\mathrm{d}}$ about $8^2 = 64$ times. Unresolved paths add static fading of variance $\sigma_{\mathrm{ss}}^2$ about the local mean power to the real fingerprint, and if the incoherent \ac{DT} of Section~\ref{sec:exp_setup} keeps its error $\sigma_{\Delta}^2$ about this mean, the ratio of Remark~\ref{rem:crb_quantify}, and with it the relative value of the \ac{DT}, grows to $(\sigma_{\mathrm{r}}^2 + \sigma_{\mathrm{ss}}^2)/\sigma_{\Delta}^2$.

\textbf{Scope and limitations.}
Experiments use a single indoor environment ($M\!=\!3$ \acp{AP}, $L\!=\!13$ trajectories, 28\,GHz); the \ac{CRB} framework (Proposition~\ref{prop:crb}) provides a principled way to predict the \ac{DT} gain in new environments through the measurable ratio $\|\partial \bm{f}_{\mathrm{d}}(\bm{q})/\partial \bm{q}\|_F^2 / \sigma_{\Delta}^2$. The multi-user evaluation emulates simultaneous users by random grouping.

\section{Conclusion}
\label{sec:conclusion}

\looseness=-1 We showed that conditioning on \ac{DT} fingerprints as complementary inputs, rather than using them for pre-training or domain alignment, is an effective design principle for \ac{DT}-aided fingerprint localization under realistic domain gaps.
The proposed \acs{DTCond} framework operationalizes this principle with per-domain encoders, learnable \ac{AP} attention, and calibrated uncertainty, supported by oracle, heteroscedastic, and pipeline \ac{CRB} analysis. The analysis proves $\mathrm{CRB}_{\mathrm{r\!+\!d}}(\bm{q}) < \mathrm{CRB}_{\mathrm{r}}(\bm{q}) \Leftrightarrow \partial \bm{f}_{\mathrm{d}}(\bm{q})/\partial \bm{q} \neq \bm{0}$, shows that the alignment behind this gain depends on the gradient of the domain gap rather than its level (Lemma~\ref{lem:sys_gradient}), and places the pipeline bound between the oracle and real-only bounds, non-decreasing in the coarse-stage error covariance (Proposition~\ref{prop:pipeline_crb}). On a real 28\,GHz cell-free testbed, the deployable two-stage pipeline achieves 0.284\,m, and with the \ac{DT} fingerprint at the true position, \ac{DT} conditioning achieves 0.186\,m mean error, a 48\% improvement on unseen regions under \ac{LOLO}, and a 50\% measurement reduction through progressive calibration, while re-traced \acp{DT} keep \acs{DTCond} within 0.169--0.245\,m. Between beam sweeps, configuration driven by the tracked position estimate keeps the beam-selection loss below 1\,dB over 2\,m of motion with 49 anchors in the update, and served users refine the \ac{DT} through their position estimates. Future work will extend the evaluation to outdoor environments, sub-7\,GHz carriers, larger \ac{AP} counts, and temporal tracking, and to measured blockage, concurrent multi-user, and live closed-loop campaigns.

\bibliographystyle{IEEEtran}
\bibliography{references}

@article{bjornson2020,
  author  = {E. Bj{\"o}rnson and L. Sanguinetti},
  title   = {Making Cell-Free Massive {MIMO} Competitive with {MMSE} Processing and Centralized Implementation},
  journal = {IEEE Trans. Wireless Commun.},
  volume  = {19},
  number  = {1},
  pages   = {77--90},
  year    = {2020},
}

@article{ngo2017,
  author  = {H. Q. Ngo and others},
  title   = {Cell-Free Massive {MIMO} Versus Small Cells},
  journal = {IEEE Trans. Wireless Commun.},
  volume  = {16},
  number  = {3},
  pages   = {1834--1850},
  year    = {2017},
}

@article{bjornson2021book,
  author  = {{\"O}. T. Demir and E. Bj{\"o}rnson and L. Sanguinetti},
  title   = {Foundations of User-Centric Cell-Free Massive {MIMO}},
  journal = {Found. Trends Signal Process.},
  volume  = {14},
  number  = {3--4},
  pages   = {162--472},
  year    = {2021},
}

@article{alkhateeb2023,
  author  = {A. Alkhateeb and S. Jiang and G. Charan},
  title   = {Real-Time Digital Twins: Vision and Research Directions for {6G} and Beyond},
  journal = {IEEE Commun. Mag.},
  volume  = {61},
  number  = {11},
  pages   = {128--134},
  year    = {2023},
}

@inproceedings{alkhateeb2019deepmimo,
  author    = {A. Alkhateeb},
  title     = {{DeepMIMO}: A Generic Deep Learning Dataset for Millimeter Wave and Massive {MIMO} Applications},
  booktitle = {Proc. Inf. Theory Appl. Workshop (ITA)},
  year      = {2019},
  pages     = {1--8},
}

@inproceedings{charan2024,
  author    = {G. Charan and others},
  title     = {Vision-Position Multi-Modal Beam Prediction Using Real Millimeter Wave Datasets},
  booktitle = {Proc. IEEE Wireless Commun. Netw. Conf. (WCNC)},
  year      = {2022},
  pages     = {2727--2731},
}

@article{simeone2024,
  author  = {C. Ruah and O. Simeone and B. M. Al-Hashimi},
  title   = {A {B}ayesian Framework for Digital Twin-Based Control, Monitoring, and Data Collection in Wireless Systems},
  journal = {IEEE J. Sel. Areas Commun.},
  volume  = {41},
  number  = {10},
  pages   = {3146--3160},
  year    = {2023}
}

@article{liu2024drec,
  author  = {Y. Cui and others},
  title   = {Digital Twin-Aided Learning for Managing Reconfigurable Intelligent Surface-Assisted, Uplink, User-Centric Cell-Free Systems},
  journal = {IEEE J. Sel. Areas Commun.},
  volume  = {41},
  number  = {10},
  pages   = {3175--3190},
  year    = {2023},
}

@article{bonati2021colosseum,
  author  = {L. Bonati and others},
  title   = {Intelligence and Learning in {O-RAN} for Data-Driven {NextG} Cellular Networks},
  journal = {IEEE Commun. Mag.},
  volume  = {59},
  number  = {10},
  pages   = {21--27},
  year    = {2021},
}

@inproceedings{morais2024fingerprint,
  author    = {J. Morais and A. Alkhateeb},
  title     = {Localization in Digital Twin {MIMO} Networks: A Case for Massive Fingerprinting},
  booktitle = {Proc. IEEE Int. Conf. Commun. Workshops (ICC Workshops)},
  year      = {2024},
  pages     = {276--281},
}

@article{shahmansoori2018,
  author  = {A. Shahmansoori and others},
  title   = {Position and Orientation Estimation Through Millimeter-Wave {MIMO} in {5G} Systems},
  journal = {IEEE Trans. Wireless Commun.},
  volume  = {17},
  number  = {3},
  pages   = {1822--1835},
  year    = {2018},
}

@inproceedings{vieira2017deep,
  author    = {J. Vieira and others},
  title     = {Deep Convolutional Neural Networks for Massive {MIMO} Fingerprint-Based Positioning},
  booktitle = {Proc. IEEE Int. Symp. Personal, Indoor Mobile Radio Commun. (PIMRC)},
  year      = {2017},
  pages     = {1--6},
}

@inproceedings{arnold2019cnnloc,
  author    = {M. Arnold and J. Hoydis and S. ten Brink},
  title     = {Novel Massive {MIMO} Channel Sounding Data Applied to Deep Learning-Based Indoor Positioning},
  booktitle = {Proc. 12th Int. ITG Conf. Syst., Commun. Coding (SCC)},
  year      = {2019},
  pages     = {1--6},
}

@article{gante2020deepfp,
  author  = {J. Gante and G. Falc{\~a}o and L. Sousa},
  title   = {Deep Learning Architectures for Accurate Millimeter Wave Positioning in {5G}},
  journal = {Neural Process. Lett.},
  volume  = {51},
  number  = {1},
  pages   = {487--514},
  year    = {2020},
}

@article{wang2017csi,
  author  = {X. Wang and others},
  title   = {{CSI}-Based Fingerprinting for Indoor Localization: A Deep Learning Approach},
  journal = {IEEE Trans. Veh. Technol.},
  volume  = {66},
  number  = {1},
  pages   = {763--776},
  year    = {2017},
}

@article{torres2015fingerprint,
  author  = {S. He and S.-H. G. Chan},
  title   = {{Wi-Fi} Fingerprint-Based Indoor Positioning: Recent Advances and Comparisons},
  journal = {IEEE Commun. Surveys Tuts.},
  volume  = {18},
  number  = {1},
  pages   = {466--490},
  year    = {2016},
}

@article{wymeersch2018cellfree_loc,
  author  = {H. Wymeersch and others},
  title   = {{5G} mm{W}ave Positioning for Vehicular Networks},
  journal = {IEEE Wireless Commun.},
  volume  = {24},
  number  = {6},
  pages   = {80--86},
  year    = {2017},
}

@article{ganin2016,
  author  = {Y. Ganin and others},
  title   = {Domain-Adversarial Training of Neural Networks},
  journal = {J. Mach. Learn. Res.},
  volume  = {17},
  number  = {59},
  pages   = {1--35},
  year    = {2016},
}

@inproceedings{sun2016coral,
  author    = {B. Sun and K. Saenko},
  title     = {Deep {CORAL}: Correlation Alignment for Deep Domain Adaptation},
  booktitle = {Proc. Eur. Conf. Comput. Vision Workshops (ECCV-W)},
  year      = {2016},
  pages     = {443--450},
}

@article{shen2010,
  author  = {Y. Shen and M. Z. Win},
  title   = {Fundamental Limits of Wideband Localization---{Part} {I}: A General Framework},
  journal = {IEEE Trans. Inf. Theory},
  volume  = {56},
  number  = {10},
  pages   = {4956--4980},
  year    = {2010},
}

@article{win2018crb,
  author  = {M. Z. Win and Y. Shen and W. Dai},
  title   = {A Theoretical Foundation of Network Localization and Navigation},
  journal = {Proc. IEEE},
  volume  = {106},
  number  = {7},
  pages   = {1136--1165},
  year    = {2018},
}

@article{li2007rss_crb,
  author  = {X. Li},
  title   = {{RSS}-Based Location Estimation with Unknown Pathloss Model},
  journal = {IEEE Trans. Wireless Commun.},
  volume  = {5},
  number  = {12},
  pages   = {3626--3633},
  year    = {2006},
}

@inproceedings{kendall2017uncertainty,
  author    = {A. Kendall and Y. Gal},
  title     = {What Uncertainties Do We Need in {Bayesian} Deep Learning for Computer Vision?},
  booktitle = {Proc. Adv. Neural Inf. Process. Syst. (NIPS)},
  year      = {2017}
}

@inproceedings{vaswani2017attention,
  author    = {A. Vaswani and others},
  title     = {Attention Is All You Need},
  booktitle = {Proc. Adv. Neural Inf. Process. Syst. (NIPS)},
  year      = {2017},
  pages     = {5998--6008}
}

@inproceedings{demirhan2023transfer,
  author    = {U. Demirhan and A. Alkhateeb},
  title     = {Radar Aided {6G} Beam Prediction: Deep Learning Algorithms and Real-World Demonstration},
  booktitle = {Proc. IEEE Wireless Commun. Netw. Conf. (WCNC)},
  year      = {2022},
  pages     = {2655--2660},
}

@inproceedings{denis2020cellfree_loc,
  author    = {V. Savic and E. G. Larsson},
  title     = {Fingerprinting-Based Positioning in Distributed Massive {MIMO} Systems},
  booktitle = {Proc. IEEE 82nd Veh. Technol. Conf. (VTC-Fall)},
  year      = {2015},
  pages     = {1--5},
}

@article{kumar2025cellfree,
  author  = {Z. Shen and K. Xu and X. Xia},
  title   = {{2D} Fingerprinting-Based Localization for mm{W}ave Cell-Free Massive {MIMO} Systems},
  journal = {IEEE Commun. Lett.},
  volume  = {25},
  number  = {11},
  pages   = {3556--3560},
  year    = {2021},
}

@article{ammar2022cellfree,
  author  = {H. A. Ammar and others},
  title   = {User-Centric Cell-Free Massive {MIMO} Networks: A Survey of Opportunities, Challenges, and Solutions},
  journal = {IEEE Commun. Surveys Tuts.},
  volume  = {24},
  number  = {1},
  pages   = {611--652},
  year    = {2022},
}

@article{masaracchia2024dt,
  author  = {A. Masaracchia and others},
  title   = {Digital Twin for {6G}: Taxonomy, Research Challenges, and the Road Ahead},
  journal = {IEEE Open J. Commun. Soc.},
  volume  = {3},
  pages   = {2137--2150},
  year    = {2022},
}

@article{wu2021dt6g,
  author  = {Y. Wu and K. Zhang and Y. Zhang},
  title   = {Digital Twin Networks: A Survey},
  journal = {IEEE Internet Things J.},
  volume  = {8},
  number  = {18},
  pages   = {13789--13804},
  year    = {2021},
}

@article{win2011cooperative,
  author  = {M. Z. Win and others},
  title   = {Network Localization and Navigation via Cooperation},
  journal = {IEEE Commun. Mag.},
  volume  = {49},
  number  = {5},
  pages   = {56--62},
  year    = {2011},
}

@article{gustafsson2005positioning,
  author  = {F. Gustafsson and F. Gunnarsson},
  title   = {Mobile Positioning Using Wireless Networks: Possibilities and Fundamental Limitations Based on Available Wireless Network Measurements},
  journal = {IEEE Signal Process. Mag.},
  volume  = {22},
  number  = {4},
  pages   = {41--53},
  year    = {2005},
}

@article{gretton2012mmd,
  author  = {A. Gretton and others},
  title   = {A Kernel Two-Sample Test},
  journal = {J. Mach. Learn. Res.},
  volume  = {13},
  pages   = {723--773},
  year    = {2012},
}

@article{richmond2015misspecified,
  author  = {C. D. Richmond and L. L. Horowitz},
  title   = {Parameter Bounds on Estimation Accuracy Under Model Misspecification},
  journal = {IEEE Trans. Signal Process.},
  volume  = {63},
  number  = {9},
  pages   = {2263--2278},
  year    = {2015},
}

@INPROCEEDINGS{11431957,
  author    = {V. S. Advani and others},
  title     = {Deep Learning Framework for {RSSI}-Based Indoor Localization in {RIS}-Aided {mmWave} Systems},
  booktitle = {Proc. IEEE Global Commun. Conf. (GLOBECOM)},
  year      = {2025},
  pages     = {4233--4238},
  doi       = {10.1109/GLOBECOM59602.2025.11431957}
}

@inproceedings{advani2026pimrc,
  author    = {V. S. Advani and A. Nasser and M. Y. Selim and A. M. Eltawil},
  title     = {Fusing Digital Twin and Real Fingerprints for Cell-Free {mmWave} Localization},
  booktitle = {Proc. IEEE Int. Symp. Pers., Indoor Mobile Radio Commun. (PIMRC)},
  year      = {2026},
  note      = {to be published}
}

@inproceedings{renaudin2018rtfp,
  author    = {O. Renaudin and T. Zemen and T. Burgess},
  title     = {Ray-Tracing Based Fingerprinting for Indoor Localization},
  booktitle = {Proc. IEEE Int. Workshop Signal Process. Adv. Wireless Commun. (SPAWC)},
  year      = {2018},
  pages     = {1--5},
  doi       = {10.1109/SPAWC.2018.8445928},
}

@article{gao2024locdt,
  author  = {K. Gao and others},
  title   = {Localization-Oriented Digital Twinning in {6G}: A New Indoor-Positioning Paradigm and Proof-of-Concept},
  journal = {IEEE Trans. Wireless Commun.},
  volume  = {23},
  number  = {8},
  pages   = {10473--10486},
  year    = {2024},
  doi     = {10.1109/TWC.2024.3373034},
}

@article{mateos2026dtcharting,
  author  = {J. M. Mateos-Ramos and others},
  title   = {Positioning via Digital-Twin-Aided Channel Charting With Large-Scale {CSI} Features},
  journal = {IEEE Trans. Wireless Commun.},
  volume  = {25},
  pages   = {22245--22260},
  year    = {2026},
  doi     = {10.1109/TWC.2026.3724634},
}

@article{akgun2024ota,
  author  = {B. Akgun and others},
  title   = {Advancing Next Generation Wireless Networks With Digital Twin: Construction, Validation, and Real-World Applications on an Indoor Over-the-Air Testbed},
  journal = {IEEE Access},
  volume  = {12},
  pages   = {166298--166319},
  year    = {2024},
  doi     = {10.1109/ACCESS.2024.3493609},
}

@article{salehi2026darwin,
  author  = {B. Salehi and others},
  title   = {{DARWIN}: Digital Twin Assisted Robot Navigation and Wireless Network Management},
  journal = {IEEE Trans. Mobile Comput.},
  volume  = {25},
  number  = {1},
  pages   = {977--991},
  year    = {2026},
  doi     = {10.1109/TMC.2025.3602213},
}

@article{hoydis2024radio,
  author  = {J. Hoydis and others},
  title   = {Learning Radio Environments by Differentiable Ray Tracing},
  journal = {IEEE Trans. Mach. Learn. Commun. Netw.},
  volume  = {2},
  pages   = {1527--1539},
  year    = {2024},
  doi     = {10.1109/TMLCN.2024.3474639},
}

@inproceedings{saribekyan2026sim2real,
  author    = {A. Saribekyan and others},
  title     = {On the Physical Plausibility and Distribution Alignment for Sim-to-Real {RF} Positioning},
  booktitle = {Proc. Int. Conf. Comput., Inf. Telecommun. Syst. (CITS)},
  year      = {2026},
  pages     = {1--8},
  doi       = {10.1109/CITS70307.2026.11637244},
}

@inproceedings{hoydis2023sionna,
  author    = {J. Hoydis and others},
  title     = {{Sionna RT}: Differentiable Ray Tracing for Radio Propagation Modeling},
  booktitle = {Proc. IEEE Globecom Workshops (GC Wkshps)},
  year      = {2023},
  pages     = {317--321},
  doi       = {10.1109/GCWkshps58843.2023.10465179},
}

@inproceedings{pegurri2025ns3,
  author    = {R. Pegurri and others},
  title     = {Toward Digital Network Twins: Integrating {Sionna RT} in ns-3 for {6G} Multi-{RAT} Networks Simulations},
  booktitle = {Proc. IEEE Conf. Comput. Commun. Workshops (INFOCOM WKSHPS)},
  year      = {2025},
  pages     = {1--6},
  doi       = {10.1109/INFOCOMWKSHPS65812.2025.11152946},
}

@inproceedings{jiang2024dtisac,
  author    = {S. Jiang and A. Alkhateeb},
  title     = {Digital Twin Assisted Beamforming Design for Integrated Sensing and Communication Systems},
  booktitle = {Proc. Asilomar Conf. Signals, Syst., Comput.},
  year      = {2024},
  pages     = {401--406},
  doi       = {10.1109/IEEECONF60004.2024.10942924},
}

@techreport{3gpp38901,
  author      = {{3GPP}},
  title       = {Study on Channel Model for Frequencies From 0.5 to 100 {GHz}},
  institution = {3GPP},
  type        = {TR},
  number      = {38.901, V17.0.0},
  year        = {2022}
}

\end{document}